\documentclass[a4paper,11pt]{article}
\pdfoutput=1 

\usepackage{jcappub}
\usepackage{tikz} 
\usepackage{caption}
\usepackage{subcaption}
\usepackage{bm}

\newcommand{\Poisson}[2] {\ensuremath{\mathcal{P}\left(#1|#2\right)}}
\newcommand{\Gaussian}[2]{\ensuremath{\mathcal{G}\left(#1|#2\right)}}

\newcommand{\myunit}{\ensuremath{\mathrm{m}^{-2}\;\mathrm{yr}^{-1}\;\mathrm{MeV}^{-1}\;\mathrm{pixel}^{-1}}}
\newcommand{\SAunit}{\ensuremath{\mathrm{cm}^{-2}~\mathrm{s}^{-1}~\mathrm{sr}^{-1}~\mathrm{MeV}^{-1}}}

\title{Modelling the flux distribution function of the
extragalactic gamma-ray background from dark matter annihilation}
\author[a]{Michael R. Feyereisen,}
\author[a]{Shin'ichiro Ando}
\author[b,c]{and Samuel K. Lee}

\affiliation[a]{GRAPPA Institute, University of Amsterdam, Science Park
904, 1098 XH Amsterdam, Netherlands}
\affiliation[b]{Princeton Center for Theoretical Science, Princeton
University, Princeton, NJ 08544, USA}
\affiliation[c]{Broad Institute, 75 Ames Street, Cambridge, MA 02142, USA}

\emailAdd{m.r.feyereisen@uva.nl}
\emailAdd{s.ando@uva.nl}
\emailAdd{slee@broadinstitute.org}

\abstract{
The one-point function (i.e., the isotropic flux distribution) is a complementary method to (anisotropic) two-point
correlations in searches for a gamma-ray dark matter annihilation
signature. Using analytical models of structure formation
and dark matter halo properties, we compute the gamma-ray flux
distribution due to annihilations in extragalactic dark matter halos, as
it would be observed by the Fermi Large Area Telescope.
Combining the central limit theorem and Monte Carlo sampling, we show that the flux distribution takes the
form of a narrow Gaussian of `diffuse' light, with an `unresolved point source' power-law tail as a result of bright halos.
We argue that this background due to dark matter
constitutes an irreducible and significant background component for
point-source annihilation searches with galaxy clusters and dwarf
spheroidal galaxies, modifying the predicted signal-to-noise ratio.
A study of astrophysical backgrounds to this signal reveals that the
shape of the total gamma-ray flux distribution is very sensitive to
the contribution of a dark matter component, allowing us to forecast
promising one-point upper limits on the annihilation cross section.
We show that by using the flux distribution at only one energy bin, one
can probe the canonical cross section required for
explaining the relic density, for dark matter of masses around tens of
GeV.}


\begin{document}
\maketitle

\section{Introduction}

The Large Area Telescope (LAT) onboard the Fermi
satellite~\cite{Fermi2009Specs} measured the energy
spectrum~\cite{Ackermann:2014usa} and angular
anisotropies~\cite{Ackermann:2012uf} of the diffuse extragalactic
background of gamma rays.
Components contributing to this background include
blazars~\cite{abdo2010fermi, Ajello:2015mfa}, star-forming and starburst
galaxies~\cite{Tamborra:2014xia}, and misaligned active
galaxies~\cite{DiMauro:2013xta}.
The combination of these sources gives reasonably good fit to the
spectral data~\cite{Ajello:2015mfa, DiMauro:2015tfa}, while the
anisotropies are consistent with the blazar component
alone~\cite{Ackermann:2012uf, Cuoco:2012yf}.
Independently of this, dark matter has emerged as the preferred
explanation of many astrophysical and cosmological features through
gravity (galactic rotation curves, $\Omega_m \gg \Omega_b$, lensing by
galaxy clusters, etc.).
If particle dark matter produces gamma rays (e.g., by self-annihilation)
as in the case of weakly interacting massive particles (WIMPs) motivated
by popular particle-physics models~\cite{Jungman:1995df, Hooper:2007qk},
then it could also contribute to this diffuse signal (in some unknown
proportion)~\cite{Fornasa:2015qua}.
Given that the known astrophysical sources yield reasonable fit to the
spectrum of the gamma-ray background, the dark matter component started
to be tightly constrained only through the spectral data
(e.g.,~\cite{Ajello:2015mfa, DiMauro:2015tfa, Ando:2015qda}).

Recently, new analysis techniques beyond the energy spectrum and angular two-point correlations were proposed and
investigated extensively.
Among them is to take cross correlations of gamma-ray data with local galaxy
catalogs~\cite{AndoCrossCorrel, Ando:2014aoa} and matter
distribution through lensing data~\cite{Camera:2012cj, Camera:2014rja}.
Although recent measurements of the cross
correlations~\cite{xia2011cross, Shirasaki:2014noa, Fornengo:2014cya,
Xia:2015wka} are consistent with the hypothesis of no dark matter signal, they yield tight constraints thereof (e.g., \cite{Regis:2015zka}).

Complementary to the studies on these two-point functions, the
one-point function (i.e. the photon-count or flux distribution) would
leverage the isotropic component of the diffuse signal.
For example, the flux distribution of Milky-Way subhalos has been used
to constrain particle dark matter properties in light of Fermi
unidentified sources~\cite{berlin2014stringent}.
The one-point function of Fermi-LAT data has
been experimentally fit to a combination of the diffuse background and
blazar-like sources by Ref.~\cite{MalyshevHogg2011}.
Theoretically, Ref.~\cite{lee2009Microhalo} studied the one-point
probability density function (PDF) of the gamma-ray flux due to
Galactic subhalos, and showed that it features power-law tail at
high-flux end (see also Refs.~\cite{Dodelson:2009ih, Baxter:2010fr}).
Understanding the one-point PDF for all the relevant sources will be
important also for possible detections of dwarfs or galaxy clusters
with gamma rays (e.g., \cite{carlson2014improving}).

In this paper, we extend the theoretical framework of
Ref.~\cite{lee2009Microhalo} to include the contribution of dark matter
annihilation in the extragalactic halos.
We model the gamma-ray flux from the population of dark matter halos
using the mass and luminosity functions predicted for the structure
formation scenario in the Universe with cold dark matter and
cosmological constant ($\Lambda$CDM).
Since there are many such halos, the total flux observed by
Fermi is predicted using `large $N$' statistical tools.
By combining the central limit theorem (CLT) in the low-flux regime and a Monte
Carlo method in the high-flux regime, we find that the differential flux distribution has a roughly Gaussian
peak, as result of the diffuse emission of a large number of very faint sources, but with a power-law high-flux tail due to the rare occurrence of an exceptionally
bright halo.
The all-sky flux in our fiducial model lies around the sensitivity limit
of the Fermi, well-below the measured gamma-ray
background~\cite{Ackermann:2014usa}, and at roughly the same level as
the dark matter flux expected from the Fornax galaxy cluster.
We find that the detectability of a dark matter signature from galaxy
clusters over the extragalactic dark matter background decreases, when
the luminosity boost due to halo substructure increases.
We also illustrate how to disentangle a dark matter signal from
astrophysical backgrounds in the presence of the photon shot noise, and
forecast the upper limit on the annihilation cross section one might
expect to obtain using the one-point function alone (although actually
performing this analysis with its due rigour is beyond the scope of this
theoretical paper). Given a fiducial model for the dark matter halo
substructure boost, the particle dark matter mass, etc., we find a
$5\sigma$ upper bound roughly a factor of two above the thermal cross
section.

This paper is organised as follows. In Sec.~\ref{sec:formalism}, we
construct the model of the flux distribution observed at the Fermi-LAT.
In Sec.~\ref{sec:model}, we detail our specific model choices, and in
Sec.~\ref{sec:results}, we present our main results, including a
sensitivity analysis of our distribution to the model choices, and a
probabilistic method for summing the fluxes from a (quite literally)
astronomically large number $\mathcal{O}(10^{22})$ of halos.
In Sec.~\ref{sec:astrodiscuss}, we discuss consequences of this study
for indirect DM searches, by comparing the predicted distribution
to the gamma-ray fluxes of galaxy clusters, dwarf spheroidals, and
blazars.
We conclude the paper in Sec.~\ref{sec:concl}.

\section{Flux probability density function: General formalism}
\label{sec:formalism}

The goal of the present work is to theoretically predict the PDF $P(F)$,
which gives the probability of observing a total gamma-ray flux $F$
arising from dark matter annihilation in extragalactic halos in a
Fermi pixel of a particular size.
In this section, we present a formalism for constructing $P(F)$ given
models for the cosmology, halo properties, and annihilation process; we
proceed in a completely general manner, postponing specific choices for
these models until Sec.~\ref{sec:model}.

We construct a Bayesian hierarchical model, to predict unknown gamma-ray
observables from well-constrained $\Lambda$CDM parameters and fitted
models of N-body simulations.
In the hierarchical Bayesian approach, uncertainties in the parameters
of probability distributions are modelled with their own distributions
(and recursively).
This allows us to systematically combine the uncertainties on physics at
widely differing scales, and thereby to perform a sensitivity analysis
of our model (Sec.~\ref{P1SA}).

\subsection{PDF for the flux from individual halos}
\label{sec:P1Fformalism}

Throughout the paper, $F$ represents the {\it differential}
flux, i.e., a number of photons received per unit area, unit time, and
unit energy range [$F(E) = d^3N_\gamma / dA dt dE$].
The PDF $P(F)$ for observing a total differential flux $F$ from all of
the halos in a pixel depends on the PDF $P_1(F)$ for observing $F$ from
any individual halo.\footnote{Throughout this paper, we denote probability
distributions by $P(\cdots)$ and distinguish them using the random
variables that they describe, along with subscripts if
necessary. Conditional and parameterised distributions are denoted as
$P(\cdot|\cdot)$. Exceptions to this convention are Poisson and Normal
distributions, denoted $\Poisson{\cdot}{\cdot}$ and
$\Gaussian{\cdot}{\cdot,\cdot}$ respectively.}
We thus proceed by first deriving the latter quantity.

Because the differential flux $F$ from an individual halo is
completely determined by its rest-frame differential luminosity $L =
d^2N_\gamma / dt dE$ and its redshift $z$, we can write
\begin{eqnarray} \label{eq:P1F-LZ}
P_1(F) &=& \int\! dL\, dz\, P(F|L,z) P(L, z) \nonumber\\
	    &=& \int\! dL\, dz\, \delta[F - F(L, z)] P(L|z) P(z)\,.
\end{eqnarray}
Here, the usual relation for the differential flux,
\begin{equation} \label{eq:differential-flux}
 F(E; L, z)=
  e^{-\tau(E, z)}\frac{(1+z)^2 L[(1+z)E]}{4\pi d_L^2(z)} \,,
\end{equation}
depends on the luminosity distance $d_L(z)$ and the pair-production
optical depth $\tau(E, z)$ for gamma-ray photons, and also accounts for
the redshift of photons emitted with rest-frame energies $E (1+z)$ to
observed energies $E$.
We can interpret $P(L|z) = dN/dL(z)$ as the redshift-dependent
halo differential-luminosity function.\footnote{Here 
and elsewhere, equalities between probability densities and number
densities is meant modulo a normalisation.}
Assuming that the halos are isotropically distributed across the
Universe, the number of halos at redshift $z$ is proportional to the
comoving volume $\delta V(z)$ of the corresponding redshift slice
$\delta z$, therefore we also have $P(z) = dN/dz \propto dV/dz$.

Alternatively, we can rewrite Eq.~\eqref{eq:P1F-LZ} in terms of the halo
mass $M$ to obtain
\begin{eqnarray} \label{eq:P1F-MZ}
P_1(F) &=& \int\! dL\, dM\, dz\, \delta[F - F(L, z)] P(L |
 M, z) P(M | z) P(z)\,,
\end{eqnarray}
where we can similarly interpret $P(M|z) = dN/dM(z)$ as the
redshift-dependent halo mass function.

In principle, the distribution $P(L | M, z)$ in Eq.~\eqref{eq:P1F-MZ}
captures the scatter in the relation between the differential luminosity
and the mass of a halo, which also depends on redshift.
This is because the halo luminosity is determined not only by the
properties of the dark matter particle and the details of the
annihilation process, but also by the density profile $\rho$ of the
halo, which usually shows scatter for any given $M$.
The halo profiles can be completely characterised by some parameters
$\bm\theta_h$ (such as $\rho_s, r_s, r_{\rm vir} \ldots$ in the case of the
NFW profile~\cite{NFW}) so that (for any given particle
dark matter model) we have $L = L(\bm\theta_h)$.
If we further assume that the distribution of halo profiles can be
described by a halo model that gives $P(\bm\theta_h | M, z)$, we can write
\begin{eqnarray} \label{eq:PL-Mz}
P(L | M, z) &=& \int\! d\bm\theta_h\, P(L| \bm\theta_h) P(\bm\theta_h
 |M, z) \nonumber\\
 &=& \int\! d{\bm \theta_h}\, \delta[L - L(\bm\theta_h)]
  P(\bm\theta_h |M, z)\,.
\end{eqnarray}
We can then use this expression to simplify Eq.~\eqref{eq:P1F-LZ},
giving
\begin{eqnarray} \label{eq:P1F-MZtheta_h}
P_1(F) &=& \int\! dM\, dz\, d\bm\theta_h\, \delta[F - F(\bm\theta_h, z)]
 P(\bm\theta_h |M, z) P(M|z) P(z)\,,
\end{eqnarray}
where the flux relation is now written in terms of $\bm\theta_h$.

In order to make the numerical calculation of
Eq.~\eqref{eq:P1F-MZtheta_h} more tractable, we shall neglect the
scatter in the distribution $P(\bm\theta_h | M, z)$ in this work.
That is, we take the distribution of the halo-profile parameters
$\bm\theta_h = \{\theta_{h,1}, \dots , \theta_{h,n}\}$ to be given by
\begin{equation} \label{eq:halo-model-delta-functions}
P(\bm\theta_h | M, z) = \prod_{i=1}^n \delta[\theta_{h,i} -
 \bar{\theta}_{h,i}(M, z)]\,,
\end{equation}
where the functions $\bar{\theta}_{h,i}(M, z)$ give the mean values for
the parameters.
With this assumption, we can perform the integrals over $\bm\theta_h$
and $M$ in Eq.~\eqref{eq:P1F-MZtheta_h}, leaving only an integral over
$z$:
\begin{equation}
P_1(F) = \int\! dz \left|\frac{\partial F}{\partial M}\right|^{-1}
 \frac{dN}{dM} \frac{dV}{dz}\,.
\label{eq:simple-P1F}
\end{equation}
Here, functions of $M$ in the integrand are evaluated at the value of
$M$ defined implicitly by the flux relation $F =
F(\bar{\bm\theta}_h(M,z), z)$ for the corresponding values of $F$
and $z$.

Eq.~\eqref{eq:simple-P1F} then gives the PDF $P_1(F)$ for the
differential flux from individual halos.
To reiterate, this is given in terms of (i) the cosmological model,
which affects the differential flux $F(\bm\theta_h, z)$, the volume
$P(z)$, and the mass function $P(M|z)$; (ii) the halo model, which gives
the mean values $\bar{\bm\theta}_{h}(M, z)$ of the halo-profile
parameters; and (iii) the optical depth and the details of the
annihilation process, which affect the normalisation of $F(\bm{\theta}_h,
z)$.
In Sec.~\ref{sec:model}, informed by observations and simulations, we
shall make specific, fiducial choices for these quantities and calculate
the resulting $P_1(F)$.
However, before doing so, we shall complete our discussion of our
general formalism by reviewing how $P_1(F)$ can be used to find
$P(F)$, the PDF of the total flux $F$ from all of the halos in a pixel.

\subsection{PDF for the total flux} \label{sec:PFformalism}

We assume \label{ptsrc} that all dark matter gamma-ray sources may be
treated as point sources (see also Sec.~\ref{extsourceperpixel} and Sec.~\ref{sec:caveats} below),
allowing us to equate the differential flux $F$ and the differential
flux per pixel.
The dark matter differential flux $F$ arriving at any given pixel of the
Fermi sky map, is the summed flux $F = \sum_i F_i$ of any number of
individual halo point sources~\cite{scheuer1957statistical}, where each
differential flux $F_i$ is an independent and identically distributed
(i.i.d.) random variable with the distribution $P_1(F)$.
The distribution of a sum of random variables is the convolution of all
the original distributions~\cite{petrov1975Sums}; Since the $F_i$ are
i.i.d., the distribution of the total differential flux per pixel is the
autoconvolution~\cite{lee2009Microhalo}
\begin{equation} \label{eq:autoconvolution}
P_k(F) = P_1(F) \star P_1(F) \star \;\cdots\; \star P_1(F) =
 \left(P_1\right)^{\star k}\,,
\end{equation}
where $k$ is the number of halos contributing to this flux. 
Since furthermore we do not know how many halos are thus stacked in a pixel,
the number $k$ of fluxes in the sum is itself a random variable.
If we assume this number $k$ of halos per pixel is Poisson-distributed
over the sky with some mean $N^\prime$, we can model the \emph{total}
differential flux per pixel as
\begin{equation}
P(F) = \int dN^\prime P(N^\prime) \sum_k \Poisson{k}{N^\prime} P_k(F)\,,
\label{eq:PFfromPk}
\end{equation}
where the uncertainties in $k$ and $N^\prime$ are marginalised away.
Since the numbers $k$ and $N^\prime$ of extragalactic halos are very
large, both $P(N^\prime)$ and $\Poisson{k}{N^\prime}$ are thin enough to
be approximated by delta functions, so that $P(F) = P_{k\approx
N^\prime}(F)$.
Thus, the only additional physical input required to compute $P(F)$
from $P_1(F)$ is $N^\prime$, which is discussed below.

\section{Model inputs of the flux probability density function}
\label{sec:model}

\subsection{Cosmological inputs}

\subsubsection{Number of halos per Fermi pixel}

Cosmology directly determines the redshift distribution of halos (via
isotropy $P(z) = dN/dz$) and their mass distribution (via the gravitational collapse
of inhomogeneities that yields $dN/dM$).
The normalisation of these number densities clearly corresponds to the
total number $N$ of halos in the Universe.
The number of halos per Fermi pixel is then
\begin{equation}
N^\prime = \frac{\Omega_{\rm pix}}{4\pi} N =  \frac{\Omega_{\rm pix}}{4\pi} \int
 \frac{dN}{dM} \frac{dV}{dz} dMdz\,,
\label{eq:Nprime}
\end{equation}
where $\Omega_{\rm pix}$ is the pixel size (expressed in units of a solid angle)
and the mass function $dN/dM$ is described in Sec.~\ref{massfunction}.
The parameters of the $\Lambda$CDM cosmology relevant to our model are
(Planck+WMAP from \cite{Planck2013CosmoParams})
\begin{equation} \label{eq:cosmoparams}
\{\Omega_b h^2,\Omega_{c} h^2,\Omega_\Lambda,H_0,\sigma_8, n_S \} =
 \{0.0221,0.120,0.685,67.3,0.83,0.96\}\,.
\end{equation}
Furthermore, the integration limits for Eq.~\eqref{eq:Nprime} (and also Eq.~\eqref{eq:simple-P1F}) are chosen as follows: we assume that our dark
matter candidate forms structures down to $10^{-6}
M_\odot$~\cite{Green:2005fa, Loeb:2005pm, Bertschinger:2006nq,
Hofmann:2001bi, Profumo:2006bv}, and allow for virialised dark matter
structures up to $10^{17} M_\odot$, 100 times more massive than galaxy
clusters.
We assume that we can measure luminosity from structures that form
between $z=5$ and $10^{-5}$, the latter of which corresponds to a
distance of roughly 45~kpc, well outside of the baryonic content of the
Milky Way in any direction.

We model the energy-dependent angular resolution of
Fermi-LAT~\cite{Fermi2009Specs} as follows:
\begin{equation}
\theta (E)  = \begin{cases}
0.8^\circ \, \left(E/\mathrm{GeV}\right)^{-0.68} & \mbox{for }
					  0.04<E/\mathrm{\mathrm{GeV}}<20 \\
0.1^\circ & \mbox{for } 20 <E/\mathrm{GeV}
\end{cases} \,,
\label{eq:FermiAngRes}
\end{equation}
such that $\theta \approx 0.8^\circ$ at our fiducial observing energy
(justified in Sec.~\ref{P1SA}) of $E=1$~GeV.

This is slightly larger than $0.6^\circ$ quoted in
\cite{Fermi2009Specs}, which is valid for normally incident photons
only.
Since we adopt this angular resolution as a size of each pixel,
$\Omega_{\rm pix} \approx \pi \theta^2(E)$, we entirely neglect the
instrumental point-spread function.
This is also justified because the number of sources per pixel is found
very large ($N^\prime \sim 7 \times 10^{21}$) and the flux is diffuse.

The largest possible mass of a point source at a given redshift
$M_\mathrm{Ext}(z)$ may be determined from the critical virial radius
$R_\mathrm{Ext} \approx d_L(z) \tan(\theta/2)$ that fits in a
pixel.
We can use this as an integration limit in Eq.~\eqref{eq:Nprime} to
cross-check our assumption (Sec.~\ref{ptsrc}) that all dark matter
gamma-ray sources are point-like.
We find less than 0.28 extended sources per pixel. This is much more than twenty
orders of magnitude smaller than the total number of
sources, but still represents a non-negligible absolute amount given the large number of pixels.\label{extsourceperpixel} We discuss this issue more thoroughly in \ref{sec:caveats}.

\subsubsection{Halo mass function \label{massfunction}}

The halo mass function, first addressed heuristically by Press and
Schechter~\cite{PressSchechter} and subsequently formalised in,
e.g., Ref.~\cite{BCEK}, is computed as
\begin{equation}
\frac{dn}{dM} = \frac{\bar{\rho}}{M} f(\nu) \frac{d\nu}{dM}\quad,\quad
 \nu = \left(\frac{\delta_c}{\sigma}\right)^2\,,
\label{eq:MassFunction}
\end{equation}
where $\delta_c = 1.69 / D(z)$ is the (linear) critical overdensity,
$D(z)$ is the linear growth factor, and $\sigma(M)$ is the rms deviation
of primordial density fluctuations, smoothed to scale $M$~\cite{BCEK}.
The functional form of $\sigma(M)$ (required to calculate $d\nu/dM$) is
determined from the literature \cite{BBKS1986} with normalisation set by the
cosmological parameter $\sigma_8$ \cite{Planck2013CosmoParams}.
The function $f(\nu)$ is derived from the excursions of these density fluctuations above a `barrier' \cite{BCEK} that encodes the physics of halo collapse
(including $\delta_c$).
For an approachable presentation of the formalism, see
Ref.~\cite{Lapi2013Statistics}.

In addition to ellipsoidal collapse, our fiducial mass function
incorporates a virialised halo's angular momentum and the cosmological
constant into its barrier $\delta_c$. It has a self-similar $f(\nu)$
well-fit by the following function (Eq. (163) in~\cite{DelPopolo:2006gn}):
\[
\nu f(\nu) \propto
\left(1+\frac{0.1218}{(a\nu)^{0.585}}+\frac{0.0079}{(a\nu)^{0.4}}\right)
\sqrt{\frac{a\nu}{2\pi}} \exp \left(-0.4019 a\nu \left[
1+\frac{0.5526}{(a\nu)^{0.585}}+\frac{0.02}{(a\nu)^{0.4}} \right]^2
\right)\,,
\] where $a = 0.707$.
The resulting mass function is similar to the more common Sheth-Tormen
parameterization \cite{SMTormen,ShethTormen} (within the resolution of
existing simulations), demonstrating that the extra physics of our
barrier have very little effect on the high-mass end of the mass
function.
However, we expect the cosmological constant to delay the formation of
large-scale structure, leaving us with a larger proportion of halos at
high redshift that are smaller than current simulations can resolve.
We find this gives roughly three times more flux than if we had used a
mass function for which the cosmological constant is ignored.

\subsection{Halo model \label{sec:halomodel}}

The differential luminosity from annihilation in a dark matter halo of
mass $M$ is given by the product of a particle physics term and an
astrophysical $J$-factor, i.e. the line-of-sight integral of the dark
matter density squared, boosted by the annihilations in halo
substructures.
The dark matter density can be parameterised using the same profiles
that fit N-body simulations well.
For an NFW profile, the $J$ factor of a point-like source can be
analytically recast as 
\begin{equation}
J=\left(1+B\right) a(c_{\rm vir}) \rho_s M_{\rm vir},
\label{eq:Jfactor}
\end{equation}
in terms of the substructure boost $B$, the virial concentration $c_{\rm
vir}$, the scale density $\rho_s$ of the profile, and the analytic
function (e.g., \cite{Ando:2009fp}) \[
a(c) = \left(1-\frac{1}{(1+c)^3}\right) \left(\ln(1+c) - \frac{c}{1+c}\right)^{-2}.
\]

The concentration parameter $c$ of an NFW halo is related to the
background density at the time that the halo forms: small mass halos are
more concentrated than high-mass halos because they form earlier
(hierarchical halo formation).
The concentration is also the link between scale parameters
$(\rho_s, r_s)$ of the halo profile and the halo's mass
content~\cite{Bullock}.\footnote{We convert between the virial and 
$(\cdots)_{200}$ conventions found in the literature using the
prescription of Ref.~\cite{HuKravtsov}. This allows us to interchangeably
convert between $c_\mathrm{vir}$ and $c_{200}$ at a given mass, and to
convert between $M_\mathrm{vir}$ and $M_{200}$ given a choice of
concentration-mass relation.}

The presence of halo substructures enhances the luminosity of the halo as a whole. Furthermore, halo substructure is expected to be denser than a host halo of the
same mass (e.g., \cite{Ando:2009fp}). This higher density entails a larger number of annihilations, further enhancing
the luminosity.
The boost factor $B$ parameterises this substructure luminosity as a
proportion of the host luminosity.
Since the $J$ factor increases as the density squared, substructure is
expected to contribute between twice to twenty times as much luminosity
as host structures of mass
$10^6<M/M_\odot<10^{13}$~\cite{SC2013Flattening}.

There are a number of N-body fitted models for NFW concentration and
substructure boost to choose from in the literature.
The models for $B$ used in this study are listed in
Table~\ref{tab:Bmodels}.

The optimistic model is based on fit to the numerical
simulations~\cite{gao2012MNRAS}, which is well motivated for mass
scales larger than the current resolution limit.
For smaller scales, on the other hand, it heavily relies on validity of
the phenomenological extrapolation.
Reference~\cite{SC2013Flattening} pointed out that such a power-law
extrapolation was unphysical, and came up with more physically motivated
model by adoping the flattening of the concentration-mass relation found
for field halos.

The boost factor for our fiducial model is given as~\cite{SC2013Flattening}:
\begin{eqnarray}
\mathrm{log}_{10} B &=& \sum_{i=0}^5 B_i \, \mathrm{log}_{10}(M_{200})^i , \label{eq:SCboost} \\
B_i &=& \left\{
-0.442,
0.0796,
-0.0025,
4.77\times 10^{-6},
4.77\times 10^{-6},
-9.69\times 10^{-8}
\right\}.
\end{eqnarray}
The underlying concentration model is very close to other
N-body-motivated models~\cite{PradaKlypin, diemer2014universal}.
The third model represents the most conservative (but
unexpected) situation in which there is no substructure boost.
In this study, we consider all the boost models in order to bracket
uncertainties on subhalos, although the fiducial model is preferred over
the others.

\begin{table}[t]
\centering
\begin{tabular}{ | l c  c |}
\hline
Boost models & [ref] & Formula\\
\hline
No boost & \nonumber & $B=0$ \\
Fiducial & \cite{SC2013Flattening} & Eq.~\eqref{eq:SCboost} \\
Optimistic & \cite{gao2012MNRAS} & $B=1.6\times 10^{-3}(M_{200}/M_\odot )^{0.39}$ \\
\hline
\end{tabular}

\caption{
Substructure boost models considered in this study, ordered from least to most optimistic in terms of annihilation signal detection prospects. The nonzero models are derived from fits to N-body simulations.
}

\label{tab:Bmodels}
\end{table}

\subsection{The gamma-ray model}

\subsubsection{Dark matter annihilation model}

The particle physics component $K(E)=\langle\sigma
v\rangle (dN/dE) /m_\chi^2$ of this model ($L(E) = J K(E)$) is
assumed here to be a standard WIMP model, with annihilations into gamma
rays via $b\bar{b}$. $\langle\sigma v\rangle$ is taken to be the thermal
cross section $3\times 10^{-26}\, \mathrm{cm}^{3}\, \mathrm{s}^{-1}$,
and the WIMP mass is taken to be $m_\chi = 85\,\mathrm{GeV}$.
Our parameterisation of the photon number per energy per annihilating
particle is \cite{bergstrom2001spectral}:
\begin{equation}
\frac{dN}{dE} = \frac{0.42\exp(-8 x)}{m_\chi(x^{1.5} +
 0.00014)} \quad,\quad x = \frac{E}{m_\chi}\, .
\label{eq:partphys}
\end{equation}

With the values quoted above, we have $K = 1.3\times 10^{-38}~\mathrm{cm}^{3}\,\mathrm{s}^{-1}\,\mathrm{MeV}^{-3}$ at 1~GeV.
Since our method is independent of any specific particle physics model,
and depends only linearly on $K$, it is trivial to rescale this study's
results to accommodate other particle physics models (as demonstrated in Sec.~\ref{sec:PC}).

\subsubsection{Gamma-ray optical depth}

By restricting our analysis to small enough redshifts and energies, we
do not need to consider photoionisation or pair production
\cite{zdziarski1989absorption}.
We can then use a very rough parameterisation \cite{bergstrom2001spectral} of gamma-ray
absorbtion:\begin{equation}
e^{-\tau(E,z)} = \exp\left[-\frac{z}{3.3} \left(\frac{E}{10 \mathrm{GeV}}\right)^{0.8}\right].
\label{eq:absorption}
\end{equation}
That this is very simplistic does not matter too much, since it is not a
very important effect at low energies and redshifts anyway --- as can be
justified quantitatively by a sensitivity analysis.

\section{Results \label{sec:results}}

In this section, we calculate $P_1(F)$ and $P(F)$ given the physical
inputs of Sec.~\ref{sec:model}.
We find that $P_1(F)$ is roughly power-law like with a log-slope of approximately $-2$ over a broad dynamic range,
with some model-dependent features to which $P(F)$ is sensitive.
More importantly, we find that $P(F)$ takes the shape of a Gaussian at
low flux, connecting smoothly with a power-law tail at high flux. This power-law tail, which has the same slope as $P_1(F)$, is the regime where the flux is dominated by a singularly bright source.

\subsection{Sensitivity analysis of $P_1(F)$ \label{P1SA}}

The probability [Eq.~\eqref{eq:P1F-MZtheta_h}] that any single halo
produces a differential flux observed with a value $F$ is then given by
marginalising away the uncertainty in its mass and redshift.
This marginalisation is clearly sensitive to the model choices we have
presented above.
We now discuss the sensitivity of $P_1(F)$ to the model choices.

\subsubsection{Dominant effects}

\begin{figure}[t]
\centering
\input{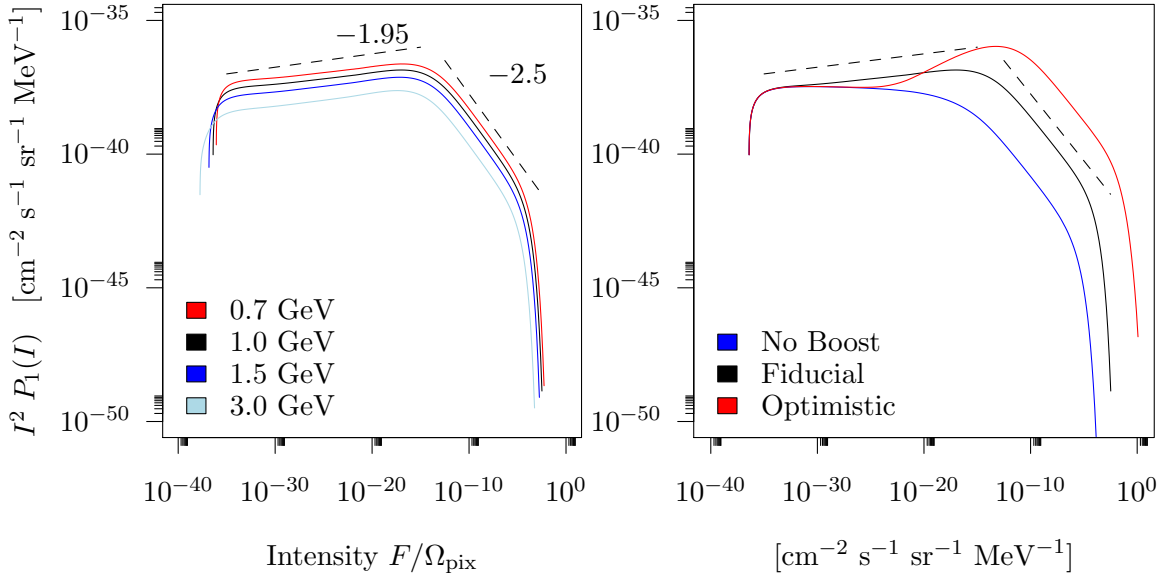}
\caption{The flux/intensity PDF for a single dark matter source, $P_1(F)$, with its dependence on photon energy (left) and boost models (right). In the right panel, the blue, black, and red curves
 represent respectively the pessimistic, fiducial, and optimistic models
 of the subhalo boost. The choice of boost model clearly and
 significantly affects the functional form of the one-point
 function. The log-slopes of the fiducial model are offset (black dashed) and quantified for convenience.
 The flux $F$ and intensity $I$ of the gamma-ray background are related via the instrument's pixel size: $F =
 I\Omega_{\rm pix}$, where $\Omega_{\rm pix} \approx 5.8\times 10^{-4}$~sr for $E =
 1$~GeV photons.
 }
\label{fig:P1SA}
\end{figure}

We find that the two most significant effects on $P_1(F)$ (from amongst
the effects we considered; see below) are the choice of observing energy
at the Fermi-LAT, and the substructure boost model $B(M)$.
These are represented in Fig.~\ref{fig:P1SA}, which
illustrates not only that $P_1(F)$ is more complex than just a power
law $P_1(F) \propto F^{-2}$ (even in the absence of a substructure
boost), but also that the substructure drastically influences the shape
of $P_1(F)$.

The choice of the energy at which we study our differential flux
represents a tradeoff between the amount of flux we expect to observe
(which decreases with energy; Fig.~\ref{fig:P1SA}), and the instrumental
capacity to actually observe it (which increases with
energy; Ref.~\cite{Fermi2009Specs}).
The observing energy of 1~GeV adopted in this study represents the lowest
energy at which we can leverage Fermi's best angular resolution and
effective area.
The instrument's energy resolution of $\Delta E / E = 9\%$ at 1~GeV is
optimum for normally incident photons~\cite{Fermi2009Specs}.
Any systematic error due to misidentified energies is therefore
minimised by this choice of energy (although we do not account for this
instrumental effect in our analysis).

Consequently, the effect of a gamma-ray optical depth (which
predominantly affects photons from distant sources) only affects the
low-flux region of $P_1(F)$.
This simply reflects the fact that at equal luminosity, distant sources
produce less flux.
At our low energy of 1~GeV, the attenuation factor of a source at $z=3$
is $e^{-\tau} \sim 0.86$: the net effect is definitely smaller than a
few percent after marginalising over the nearby halos, for which the
attenuation is truly negligible.
A better parameterisation than \cite{bergstrom2001spectral} was
therefore not deemed necessary for this exploratory analysis.

Since our differential flux is roughly $F \sim L/d_L^{2}$, there is an
extent to which `bright' halos are massive and nearby, while `faint'
halos are light and distant.
Therefore the largest modeling factor in the high-flux tail is the
choice of boost model.
This is reflected in the right panel of Fig.~\ref{fig:P1SA}: the
optimistic model~\cite{gao2012MNRAS} gives very large high-flux tail, in
comparison to the fiducial model.
The mass scale at which the fiducial and optimistic boost models
intersect ($\sim 10^{9} M_\odot$~\cite{SC2013Flattening}) is present in
the flux distribution.
We find that $P_1(F)$ is less sensitive to other modeling choices.

\subsubsection{Subleading effects}

We have used various analytical models for the power spectrum transfer
functions $T(k)$ \cite{eisenstein1998baryonic, efstathiou1986isocurvature}, and for $c_\mathrm{vir}$ fits to N-body
simulations~\cite{SC2013Flattening, diemer2014universal, PradaKlypin}.
The resulting distributions $P_1(F)$ were found to be robust to changes
in these inputs.
Indeed, the function $a(c)$ in the $J$ factor [Eq.~\eqref{eq:Jfactor}]
is very smoothly decreasing for $c>1$, so the choice of concentration
model does not influence the final result much.
Similarly, $\sigma(M)$ only changes by about an order of magnitude over
many orders of magnitude of halo masses, so varying models of $T(k)$ (or even using an N-body fit \cite{klypin2011dark})
does not significantly change $P_1(F)$ either.

Using such a simple, `self-similar' mass function, is not without
shortcomings: these Markov-process models do not follow individual halo histories,
and cannot account for dynamical effects such as dynamical friction
or tidal stripping. This, amongst other concerns,
underlines the danger of our uncontrolled extrapolation of the mass
function down to halo masses $10^{-6}M_\odot$.
Nevertheless, mass function fits to N-body simulations are often very
good; the fits even favour ellipsoidal collapse
models~\cite{SMTormen,ShethTormen} (which we adopt in this study) over
the spherical collapse models of the seminal papers.
The ellipsoidal collapse model is accurate to a few \% over the large halo mass
range (while spherical collapse~\cite{PressSchechter} underestimates the
number of FOF halos \cite{warren2006precision}) and gives an excellent
fit in the range of $10^{5}$--$10^{9} M_\odot$ \cite{colin2004dwarf}.
Ellipsoidal collapse is also suitable for halos as small as $10^3
M_\odot$ and as early as $z=15$ \cite{sasaki2014Statistical}.
We studied dependence of the mass function on $\Lambda$, in addition to
ellipsoidal collapse \cite{DelPopolo:2006gn}.
Without significantly changing $P_1(F)$,
we find that the delayed structure formation gives a roughly three times
larger total flux (by increasing the number of halos $N$) than if we had used a mass function for which the
cosmological constant is ignored \cite{SMTormen,ShethTormen}.


We will need, when computing the total $P(F)$, to compute the first few moments of $P_1(F)$ at an intermediate step:
\begin{eqnarray}
 \mathbb{E}_{P_1(F)} &=& \int dF^\prime P_1(F^\prime) F^\prime\,,
  \label{eq:F mean}\\
\mathbb{V}_{P_1(F)} &=& \int dF^\prime P_1(F^\prime) (F^\prime - \mathbb{E}_{P_1(F)})^2\,.
\label{eq:deltamoments}
\end{eqnarray} After multiplying Eq.~\eqref{eq:F mean} by the mean number of halos
[Eq.~\eqref{eq:Nprime}] and dividing by the pixel size, one obtains the
mean intensity of the gamma-ray background from dark matter
annihilation~\cite{ando2005EGB}.
Similarly the variance [Eq.~\eqref{eq:deltamoments}] is related to the
angular power spectrum after similar corrections~\cite{Ando:2005xg}.

Since Eq.~\eqref{eq:simple-P1F} entails an integration over redshift for
each $F$, and since we find $P_1(F)$ to be relatively smooth, we
calculate 250 logarithmically equidistant points over the $\sim$40
orders of magnitude supporting the distribution.
In order to obtain robust estimate of the moments, we further sample
250 points within the four orders of magnitude nearest to the maximum
estimated from the low-resolution sampling.

Sources of uncertainty not studied include the cosmological parameters
\eqref{eq:cosmoparams}, instrumental effects related to Fermi-LAT, and
the assumed NFW profile with exactly determined parameters $\bar{\bm\theta}
(M,z)$.
Of these, the largest quantifiable uncertainty is the scatter about the concentration
$c_\mathrm{200}$, estimated at 15\% \cite{diemer2014universal}.

\subsection{Computing $P(F)$}

\subsubsection{Monte Carlo method combined with the central limit
   theorem\label{sec:CLTMC}}

Once $P_1(F)$ and the number of halos per pixel $N^\prime$ is specified,
the calculation of $P(F)$ requires no additional physical assumptions.
However, the large number of halos per pixel $k \approx N^\prime$ makes
any exact calculation of the autoconvolution
[Eq.~\eqref{eq:autoconvolution}] prohibitively expensive, even using
Fourier methods~\cite{scheuer1957statistical,haines1988logarithmic,lee2009Microhalo}.
One might attempt to use the central limit theorem (CLT) to approximate
$P_k(F)$ by a Gaussian.
But, although the CLT guarantees convergence in distribution for
$k\rightarrow \infty$, at finite $k$ there will be deviations from a
Normal distribution, particularly in the
tails~\cite{petrov1995LimitTheorems}.
This is especially true for power-law-like distributions such as
$P_1(F)$, since the stable distributions of sums of power laws may be
non-Gaussian~\cite{petrov1975Sums}.
See also Appendix~\ref{CLTFAIL}.

\begin{figure}[t]
\centering
\input{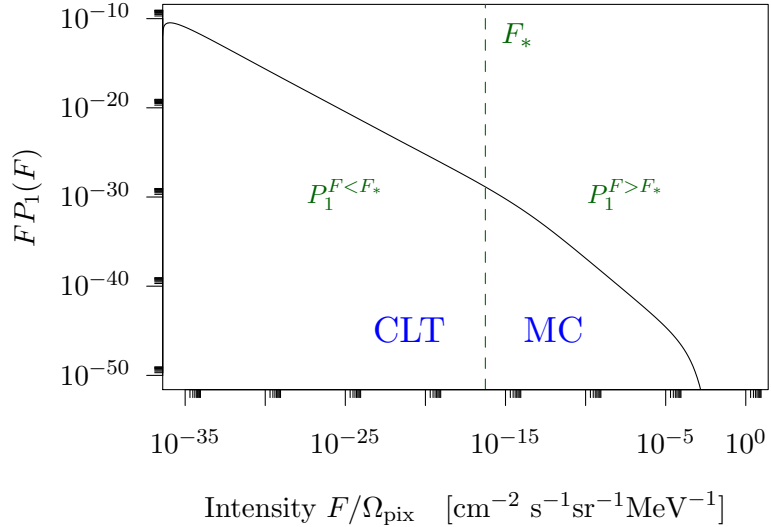}
\caption{Schematic of the $F_*$ cutoff of $P_1(F)$ into high/low
 flux. In our computation of the full $P(F)$, the central limit theorem is used to combine the fluxes from the many sources fainter than
 $F_*$, that follow a distribution $P_1^{F<F_*}(F)$. Monte Carlo is used above this cutoff to combine the halo fluxes drawn from $P_1^{F>F_*}(F)$.}
\label{fig:Fstarscheme}
\end{figure}

For the purpose of solving the autoconvolution
[Eq.~\eqref{eq:autoconvolution}], we can choose to split $P_1(F)$ into
two physically interesting contributions: the multitude of low-mass,
faint halos contributing less than some cutoff flux $F_*$; and the rare,
large point sources brighter than $F_*$.
This scheme is illustrated in Fig.~\ref{fig:Fstarscheme}.
The former will contribute an isotropic background with Gaussian
statistics, while the latter will add a power-law contribution to the
high-flux tail, which smoothly matches with $P_1(F)$ at high fluxes
(where flux is dominated by a single source).
The noise at high flux in our Monte Carlo generated $P(F)$ can therefore
be confidently ignored as a numerical artefact
(cf. Fig.~\ref{fig:Matching} in Appendix \ref{methodappendix}).

Despite the large value of $k$, our cutoff $F_*$ may be chosen such that
only a few thousand of these high-flux sources remain in each Fermi
pixel.
We can model the contribution of these sources by Monte Carlo, drawing
number of these rare sources from a Poisson distribution and their flux
from $P_1(F)$.
We obtain finally the flux from $k$ halos as the sum of these two
contributions; The flux distribution from $k$ sources is given by the
following convolution:
\begin{equation}
P_{k}(F) =\mathcal{G}_\mathrm{CLT}^{F<F_*}(F) \star P_{\mathrm{MC}}^{F>F_*}(F)~.
\label{eq:CLTconvMC}
\end{equation}

We note that when taking the convolution of our Monte Carlo result with
our faint-halo Gaussian, we care mostly about the peak of the Gaussian
(since the peak is the largest contribution to the convolution
integral).
The CLT thus offers a suitable approximation of the $P_1^{F<F_*}(F)$
autoconvolution for this purpose, despite deviations from a Normal
distribution in tails.
In fact, for practical purposes \emph{only} the peak matters and the
convolution, Eq.~\eqref{eq:CLTconvMC}, is a trivial shift of the
Monte Carlo result to higher fluxes.

For a more detailed derivation of these results, see
Appendix~\ref{methodappendix}.

\subsubsection{Flux distribution and instrumental sensitivity}

\begin{figure}[t]
\centering
\input{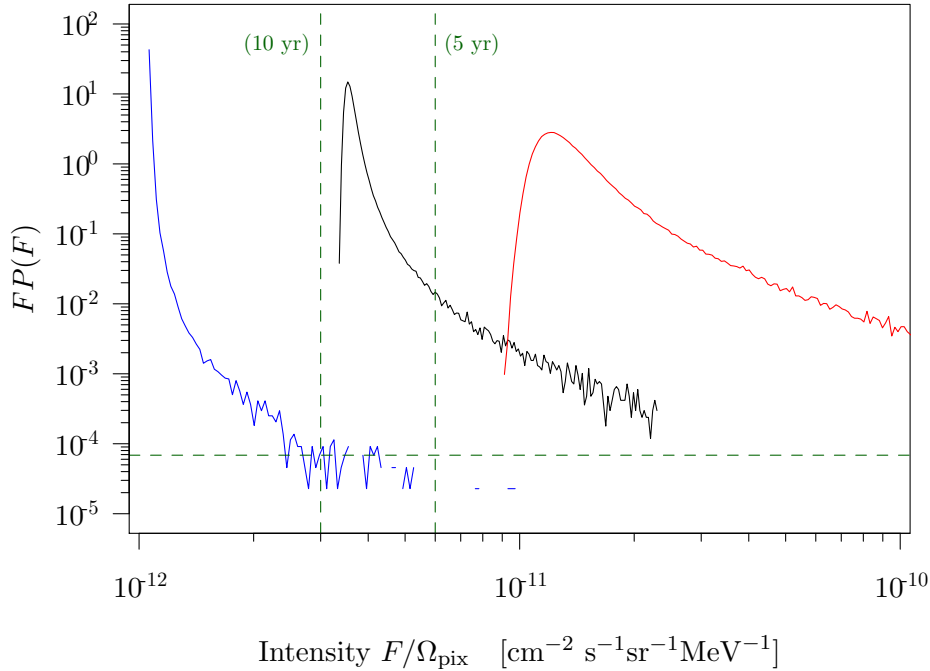}
\caption{The flux PDF $P(F)$ per pixel. The blue, black, and red curves
 represent respectively the pessimistic, fiducial, and optimistic models of the subhalo boost.
 Instrumental responses of Fermi-LAT on detecting $P(F)$
 are schematically shown. Vertical lines represent a flux corresponding
 to a single, one GeV photon per pixel, over the course of a mission of duration 5 (10) years. The Horizontal line schematises the
 angular resolution limit [Eq.~\eqref{eq:angresfmax}] at 1~GeV.}
\label{ShinFig}
\end{figure}

In Fig.~\ref{ShinFig}, we show the flux PDF $P(F)$ for the three subhalo boost models.
As discussed above, the distributions are well represented by the
`diffuse' component of nearly Gaussian with a power-law tail at high-flux
regime. The mean of this distribution [Eq.~\eqref{eq:F mean}] corresponds to $\sim$10$^{-12}\, \SAunit$ of the diffuse gamma-ray background, with an order-of-magnitude level dependence on the model (see Table~\ref{tab:meanmode}).
This is the value used as the mean intensity in the literature, in
order to constrain the dark matter annihilation cross section from the
comparison with the spectral data~\cite{Ackermann:2015tah}.
However, our PDF analysis shows that the distribution is skewed, such that the mean is not the most likely value to be observed in any given pixel: the mode is typically lower than the mean, by a boost-dependent factor of the order of a few percent, again summarised in Table~\ref{tab:meanmode}.
If one instead uses these most likely values, then the upper limits on
annihilation cross-section will accordingly remain relatively stable: existing upper limits are thereby relatively immune to the non-Gaussianity of the $P(F)$ tails. However, for accurate results, one has to perform the data analysis by
taking into account the full shape of $P(F)$.

\begin{table}[t]
\centering

\begin{tabular}{ | l | c  c | c c | c |}
\hline
Boost model & Mean & Most Likely & Difference & Ratio & EGB fraction\\
\hline
No boost & 1.0 & 1.0 & 0.0 & $\times$ 1.0 & 0.2\% \\
Fiducial & 3.68 & 3.52 &  0.16 & $\times$ 1.05 & 0.6\% \\
Optimistic & 15.2 & 11.9 & 3.3 & $\times$1.3 & 2.5\% \\
\hline
\end{tabular}
\caption{Mean and most likely extragalactic dark matter annihilation
 intensities as a function of the substructure boost model, in units of
 $10^{-12}\,\SAunit$. The difference between these quantities is percent-level, securing existing constraints on particle dark matter properties against the non-Gaussianity of $P(F)$. We also provide, for interest, the value of this mean contribution of the DM as a fraction of the unresolved EGB at 1~GeV \cite{Abdo2010spectrum}.}
\label{tab:meanmode}
\end{table}

Before contrasting our dark matter signature to known and
well-observed astrophysical sources such as galaxy clusters and blazars
in the next section, we briefly touch on whether Fermi is sensitive
enough to see it at all.
A flux of a single, GeV photons per pixel, over a 5 year mission with
LAT's effective area of $0.9~ \mathrm{m}^2$ and a field of view of 1/5 of
the sky, corresponds to a differential flux of $6\times10^{-12} \, \SAunit$. The
bulk of the one-point function $P(F)$, with its peak of $3.5\times
10^{-12}\,\SAunit$, lies just below this sensitivity limit.
See the vertical dashed lines in Fig.~\ref{ShinFig}, for the sensitivity
curves for 5-year and 10-year Fermi exposure.

There will be a small
fraction of the pixels that register photons from the high-flux tail.
Since the high-flux power-law tail is characterised by $P_1(F)$ (see Fig.~\ref{fig:Matching} in Appendix~\ref{methodappendix}), this is to some
extent equivalent to computing ``the probability of seeing a dark matter
point source.''
The finite angular resolution of the LAT limits the number of pixels in
which we can look for these bright outliers: we
argue in Appendix \ref{sec:Binomial} that a maximum flux $F_{\rm max}$, the brightest flux one might expect in any
pixel, is given by
\begin{equation}
  F_{\rm max} P(F_{\rm max}) \approx \frac{1.5}{N_{\rm pix}} \,,
  \label{eq:angresfmax}
\end{equation}
where $N_{\rm pix} = 4\pi / (\pi \theta^2) = 2.18\times10^5$ is the number of
pixels. This equation simply states that we are unlikely to probe the flux regime of
$P(F)$, where odds are worse than $1/N_\mathrm{pix}$.
In Fig.~\ref{ShinFig}, we show $F_{\rm max}P(F_{\rm max})$ as a horizontal dashed
line, and this confirms that one is able to probe this high-flux
tail with Fermi-LAT's very good angular resolution.

Since the fiducial model lies on the edge of detectability with Fermi,
it would not substantially contribute to the observed extragalactic
background with the fiducial choices of dark matter parameters; this is
consistent with what is found in the literature (e.g.,
\cite{ando2005EGB}).
Of course, the particle physics parameter space
[Eq.~\eqref{eq:partphys}] may still allow such contributions for
e.g. larger values of $\langle \sigma v \rangle$ than the thermal cross section: in Sec.~\ref{sec:PC} we will vary $\langle \sigma v \rangle$ slightly above the fiducial value, and therefore expect a flux of a few photons per pixel over the course of the 5-year mission.
Similarly, with an optimistic boost model, the dark matter may
contribute a few photons despite a using the canonical particle physics model
(Fig.~\ref{ShinFig}).

\section{Discussion\label{sec:astrodiscuss}}

\subsection{Searches for clusters and dwarf spheroidal galaxies}

\begin{figure}[t]
\centering
\input{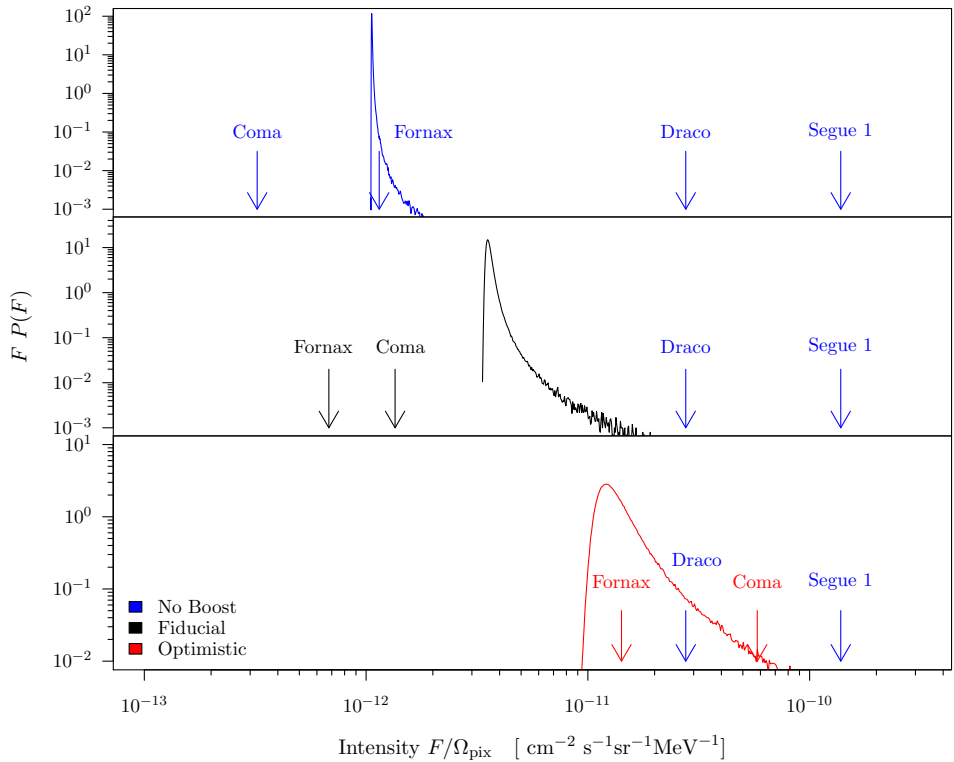}
\caption{Brightnesses of promising clusters and dwarf spheroidal
 galaxies superposed on the extragalactic dark matter annihilation gamma-ray background. The color code is the same as for previous figures. We assume that dwarf spheroidals have no substructure
 boost. The fiducial model
 does not favour indirect searches with clusters. The inversion of predictions for Coma and Fornax between top
 and bottom panels accounts for source extension, as explained in the
 main text.}
\label{fig:clusters}
\end{figure}

This diffuse emission due to dark matter annihilation characterised by $P(F)$ acts as a background to dark matter point source searches in galaxy clusters and
dwarf spheroidal galaxies. The situation is summarised in Fig.~\ref{fig:clusters}: The flux from promising candidate sources is superposed onto the extragalactic dark matter one-point function, for the three boost models presented in Table~\ref{tab:Bmodels}. A discussion of astrophysical backgrounds is postponed until Sec.~\ref{sec:unresolvedblazar}: indeed, it is unlikely to find by coincidence a bright blazar in the same pixel as a drawf galaxy.

\subsubsection{Clusters of galaxies}

Our candidate sources in this category are the Fornax and Coma clusters \cite{ando2012Fornax, ZandanelComa}. These targets are among those with the most promising $J$-factors, so that they represent the most optimistic prospects of an annihilation signal detection (or alternatively the strongest constraints on such a signal).

The substructure of a halo lies predominantly outside the scale radius
$r_{s}=r_\mathrm{vir}/c_\mathrm{vir}$~\cite{arad2004phase, gao2012MNRAS}.
In the absence of substructure boost, in contrast, almost all the
luminosity is concentrated within this scale radius and our clusters are
well approximated as point sources for Fermi-LAT:
\begin{equation}
\theta_{B=0}=\arctan\left(\frac{r_{s}}{d_A} \right) \leq
 \theta_\mathrm{pix} \,,
\end{equation}
where the fiducial concentration model $c_\mathrm{vir}(M,z)$ is used,
and $d_A$ is the angular diameter distance.

If the boost is significant, then substructure outside the scale radius
of these clusters contributes to the luminosity: light comes from the
entire virial radius, and the flux from the halo (treated as an extended
source) is diluted between many adjacent pixels.
The boost factor gives the fractions of luminosity from the host and its
substructure, so a rough estimate of the cluster radius we should
convert into an angular extension is
\begin{equation}
R = \frac{r_s + B r_{\rm vir}}{1+B}\,.
\end{equation}
We clearly recover $R \sim r_s$ when the contribution from substructure
is negligible ($B\ll1$), and $R \sim r_\mathrm{vir}$ when $B  \gg 1$.
The angle $\arctan(R/d_A)$ then determines the number of pixels over which the flux is averaged into an intensity. This corresponds to flux dilutions over roughly 10 pixels for Coma and
60 pixels for Fornax, explaining why intensities from Coma and Fornax
appear inverted in lower two panels of Fig.~\ref{fig:clusters}: the
total flux increases when considering substructure, but flux \emph{per
solid angle} decreases more for Fornax than for Coma.
The fact that the intensity from Fornax appears to decrease from
the top panel to the lower panels of Fig.~\ref{fig:clusters} is then just a manifestation of the difference
between seeing Fornax as a point source in the top panel or as an extended
source in the lower panels.

For the optimistic boost model, Coma stands out in the tail of $P(F)$,
while Fornax is only barely more visible than if it (pessimistically)
had no substructure.
Although our treatment of source extension is somewhat naive, the
diffuse gamma-ray background would be a limiting factor in cluster
analysis for the fiducial boost model, even if the effects of extension
were favourably revised by a factor of three or four (see middle panel
of Fig. \ref{fig:clusters}).

The intrinsically poor signal-to-noise in cluster searches is also
independent of the annihilation cross section or mass: changing these
particle physics parameters would not change the signal-to-noise ratio,
since both the target cluster and the gamma-ray background are rescaled
by the same factor.

\subsubsection{Dwarf spheroidal galaxies}

Dwarf spheroidal galaxies are another strategic choice for gamma-ray
point-source searches, to which dark matter annihilations in the Milky
Way substructures constitute a known background
\cite{carlson2014improving}.
We now discuss the background due to extragalactic sources.

We consider Draco and Segue 1, with $J$-factors from Ref.~\cite{ackermann2011constraining-dark-matter} since we do not expect our virialised halo model
(Sec.~\ref{sec:halomodel}) to apply to them. Again, these sources have larger
$J$-factors than other known dwarfs, and should thereby set the strongest constraints on non-detection.
These sources lie well above the isotropic background component, even if
we assume no relevant substructure boost for the dwarfs ($B=0$).
Consequently, the more substructure boost there is in extragalactic dark
matter halos, the worse the signal-to-noise for dwarfs will be. Since we know the full distribution $P(F)$ for the extragalactic background, the $p$-value of an excess signal at 1 GeV (due to a dwarf spheroidal) can readily be estimated. Even though the mean intensities yield poor signal-to-noise ratios, a pixel as bright as Segue 1 would be relatively uncommon (though not absent from the Fermi skymap, see Fig.~\ref{ShinFig}).

\subsection{Astrophysical backgrounds: Blazars and other components}

In this section we consider a number of known astrophysical backgrounds
that would mask the signal of our fiducial model.

\subsubsection{The unresolved blazar flux distribution \label{sec:unresolvedblazar}}

Following the parameterisation of the blazar source count from
Ref.~\cite{abdo2010fermi}, we assume the following power-law for
$P_1(F)$ of unresolved blazars:
\begin{eqnarray}
\frac{d^2N}{dSd\Omega} &=& 1.4\times 10^{-7}\, \label{eq:dNdS}
\left(\frac{S}{\mathrm{cm^{-2}\,s^{-1}}}\right)^{-1.64}~\mathrm{cm}^{2}~\mathrm{s}~\mathrm{deg^{-2}}\,, \\
P_1(F) &=& 2.5 \times 10^{-10}\, \left(\frac{F}{\mathrm{cm^{-2}\,s^{-1}\,MeV^{-1}}}\right)^{-1.64}~\mathrm{cm}^{2}~\mathrm{s}~\mathrm{MeV}
\, ,
\end{eqnarray}
where $S$ is the gamma-ray flux integrated above 100~MeV. We extrapolate this relation down to a lower flux limit of $S \ge 0.36\times 10^{-10}
~ \mathrm{cm}^{-2}\,\mathrm{s}^{-1}$, and perturb around this fiducial value\footnote{This value is ten times smaller than the `medium band' from Ref.~\cite{abdo2010fermi}, corresponding to the faintest \emph{observed} source out of the mix of FSRQs and BL Lacs that contributes at 1 GeV.} to test the model's sensitivity to this extrapolation (see Table \ref{tab:blaz}). We also enforce an upper limit of $S \le 2 \times 10^{-8} ~\mathrm{cm}^{-2}\,\mathrm{s}^{-1}
$, in order to maintain a low blazar detection efficiency
\cite{abdo2010fermi} without altering the power-law form of the blazar
$P_1(F)$. Assuming an $E^{-2.4}$ spectrum, the corresponding range in
differential flux at 1~GeV is then $2.0
\times 10^{-15} \le F / (\mathrm{cm}^{-2}\,\mathrm{s}^{-1}\,\mathrm{MeV}^{-1}) \le 1.1 \times 10^{-12}$.

\begin{figure}[t]
\centering
\input{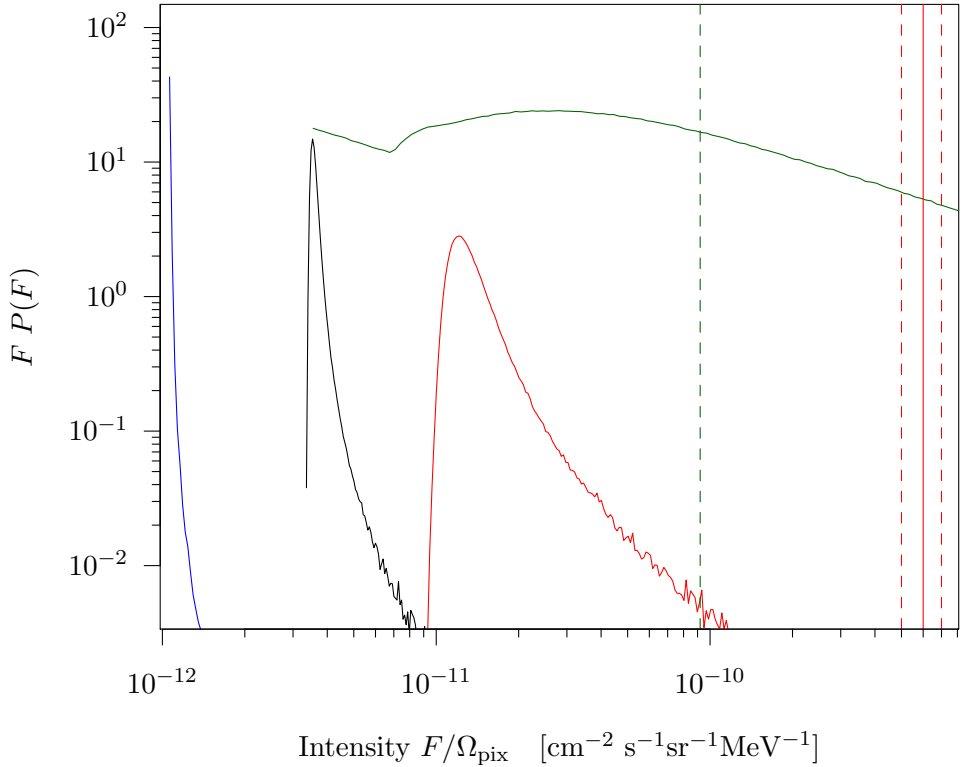}
\caption{One-point function $P(F)$ for the three dark matter models (with boosts color-coded as previously), alongside the $P(F)$ of the diffuse
 contribution of blazars (green). The dashed
 red band represents the measurement of the unresolved EGB from the Fermi data at 1~GeV \cite{Abdo2010spectrum},
 while the dashed green line is the mean of the blazar PDF.}
\label{fig:Blazars}
\end{figure}

\begin{table}[t]
\centering

\begin{tabular}{ | c c | c c |}
\hline
$S_{min}$ & $N_{blz/pix}$ & $\langle F \rangle / \Omega_{pix}$ & (\% EGB) \\
\hline
0.72 & 1.25 & 0.89 & (14.8\%) \\
0.36 & 1.97 & 0.921 & (15.3\%) \\
0.18 & 3.08 & 0.946 & (15.8\%) \\
\hline
\end{tabular}

\caption{Sensitivity summary of our unresolved blazar model (with fiducial values in the central row). The first two columns pertain to Eqn.~\eqref{eq:dNdS}: $S_{min}$ (in units of $10^{-10}
~ \mathrm{cm}^{-2}\,\mathrm{s}^{-1}$) is the lowest flux to which we extrapolate the source count distribution, from which a number of (faint, unresolved) blazars per pixel may be derived. The next two columns summarise the corresponding mean intensity of the blazar $P(F)$, both as an absolute value in units of $10^{-10}$ \SAunit, and as a proportion of the unresolved EGB \cite{Abdo2010spectrum}.}
\label{tab:blaz}
\end{table}

Using $d^2N/dSd\Omega$, we find $\sim 2.0$ faint blazars per pixel, and draw the
number of sources per pixel from a Poisson distribution in order to
predict $P(F)$ by Monte Carlo simulation.
In practice, we split the blazar $P(F)$ into two contributions: the delta-function of zero sources per pixel (treated analytically) and the contribution of more than
zero sources per pixel (Monte Carlo):
\begin{equation}
 P_\mathrm{Blazar}(F) = \sum_{k=0}^\infty \Poisson{k}{2.0} P_k(F) = e^{-2.0} \delta(F) +  \sum_{k=1}^\infty \Poisson{k}{2.0} P_k(F)\,.
 \label{eq:blazarmixture}
\end{equation} The (nonzero) blazar flux distribution is plotted in Fig.~\ref{fig:Blazars}. Like the flux distribution for the dark matter, it has a non-negligible skew and approximates the single-source distribution at high fluxes. $P_\mathrm{Blazar}(F)$ also follows the single-source powerlaw between the minimal intensity of a single source $I_{min}=F_{min}/\Omega_\mathrm{pix}$ and the minimal intensity of two sources, with a discontinuous derivative at this threshold. We see from Fig.~\ref{fig:Blazars} and Table~\ref{tab:blaz} that unresolved blazars contribute roughly $15\%$ of the unresolved EGB at 1~GeV, reproducing Ref.~\cite{abdo2010fermi}: even a single blazar is at least as bright as the entire dark matter contribution of our fiducial model, and the mean blazar flux is between one and two orders of magnitude brighter than the dark matter, depending on the boost model. Fortunately, using the entire distributions (instead of just their means) allows the thin, peaked dark matter to be statistically extracted from the broad, powerlawlike blazars (see Sec.~\ref{sec:PC}).

\subsubsection{The isotropic background}

There are a number of other backgrounds to consider, such as cosmic rays
or starburst galaxies. Since we expect all contributions of the EGB to add up to the
experimentally observed flux, we must convolve the distributions above
with the $P(F)$ of these other isotropic backgrounds. However, a complete model of these backgrounds is beyond the scope of this exploratory analysis. The one-point function of these other isotropic components is (for convenience) assumed Gaussian with a small, arbitrarily chosen, but non-negligible variance ($\mu/\sigma \sim 10^3$) and mean determined by
requiring that the mean intensity due to $P_\mathrm{EGB}(F)$ (after adding the dark matter and blazar components) be equal to the experimentally determined value $6\times 10^{-10} \;
\SAunit$ \cite{Abdo2010spectrum}. This Gaussian could be thought of as the central limit theorem approximation to the flux distribution of a large number of sources, since $\mu/\sigma$ scales like $\sqrt{N}$.

We can also consider not convolving a dark matter component into the gamma-ray
background, to study (in Sec.~\ref{sec:PC}) the distinguishability of an
alternate hypothesis model with dark matter, $P_\mathrm{Alt}(F)$, from a
null hypothesis model without dark matter $P_\mathrm{Null}(F)$ (Both of these flux distributions are represented in Fig.~\ref{fig:EGBnoDM}). We then have
\begin{equation}
P_\mathrm{EGB}(F) = P_\mathrm{Blazar}(F) \star
 \mathcal{G}_\mathrm{rest}(F)~~\star~ P_\mathrm{DM}(F)\,,
\label{eq:EGBPF}
\end{equation} which can be solved using the blazar $P(F)$ model [Eq. \eqref{eq:blazarmixture}] above, as \begin{equation}
P_\mathrm{EGB}(F) = \left[e^{-2.0} \delta(F) + (1-e^{-2.0}) P_\mathrm{Blazar}^\mathrm{MC}(F) \right] \star \mathcal{G}_\mathrm{rest}(F) ~~\star~ P_\mathrm{DM}(F)\,,
\end{equation} where the analytical factor of $(1-e^{-2.0})$ captures the normalisation of the Monte Carlo. It shall be crucial in what follows that the mean of $P_\mathrm{EGB}(F)$ has been fixed as an experimental input, rather than allowed to vary freely as a model parameter.

\subsubsection{The combined astrophysical backgrounds}

\begin{figure}[t]
\centering
\input{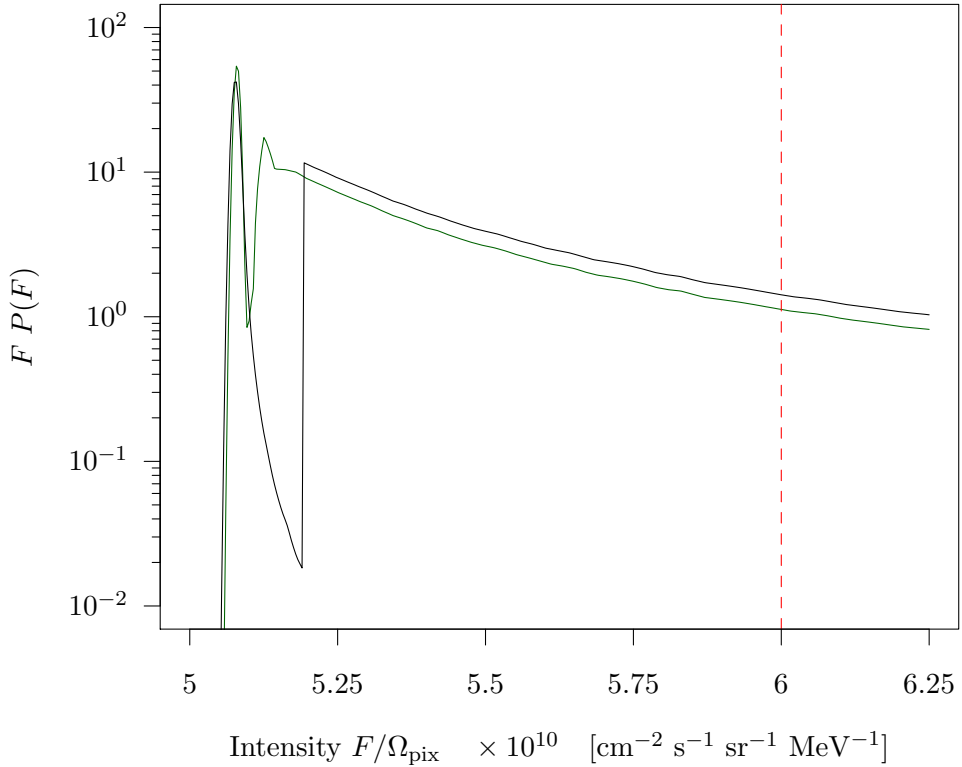}
\caption{Predicted flux distribution $P_\mathrm{EGB}(F)$ of the extragalactic gamma-ray background, with (black) and without (green) a contribution from dark matter annihilations. The distributions have two peaks, based on whether or not a blazar is present in the associated pixel. The mean EGB derived from Fermi \cite{Abdo2010spectrum} is represented by the vertical line (red, dashed). A cross-section twice the canonical value was used to visually enhance the differences between these distributions.}

\label{fig:EGBnoDM}

\end{figure}

The purpose of this section is to describe the influence of modelling choices on the shape of the flux distribution of the total unresolved EGB, as represented in Fig.~\ref{fig:EGBnoDM}.

The thin peaks and power-law-tailed peaks at low
and high flux in Fig.~\ref{fig:EGBnoDM}, correspond respectively to whether or not a blazar is present in the
associated pixel. The relative peak heights are determined (to a first approximation) by the number of unresolved blazars per pixel, via the mixture coefficient
$e^{-2.0}$. Broadening the peaks (by convolving the astrophysical components with a dark matter component) introduces a correction to this determination of height. The location of the mean of the low-flux peak is set via (i) Eqn.~\eqref{eq:EGBPF}, (ii) the (assumed) mean of the blazar flux distribution (cf. Table \ref{tab:blaz}), and (iii) the experimentally determined mean of the total flux distribution \cite{Abdo2010spectrum}. The difference between the dark matter distribution's mean and most probable fluxes (cf. Table \ref{tab:meanmode}) introduces an additional percent-level shift between the locations of the low-flux peaks with and without dark matter.

The width and depth of the `gap' between the high and low flux peaks is related, in the absence of dark matter, to the lower flux
limit on blazars assumed in Sec.~\ref{sec:unresolvedblazar}. This relation relies on the interplay between two effects: on one hand, extrapolating to lower fluxes increases the mean blazar contribution, shifting the location of the low-flux peak by an equal and opposite amount to even lower fluxes and widening the gap. On the other hand, extrapolating the faintest blazar contribution to lower fluxes also fills the gap with the fluxes from these faint blazars. The latter effect more than compensates for the former, such that overall the gap closes as it shifts to lower fluxes with an increasing blazar contribution.

However, the contribution of the dark matter component to this gap is just as dramatic. Since the distribution is skewed, it broadens both peaks preferentially to higher fluxes; however, the location of the low-flux peak is fixed by the constraint imposed on the mean of $P_\mathrm{EGB}(F)$, while the high-flux peak is free to shift under this broadening. This widens the gap with increasing brightness of the dark matter annihilation signal.

In addition to these shifts, the slope of the flux-tail due to dark matter annihilation ($-2.5$) is quite different from the one for blazars ($-1.64$); if the former component gives a
significant contribution to the gamma-ray background, this will deform the shape of the total distribution's tail. The effect of changing the variance of the isotropic Gaussian, being of less interest than actually modelling these components, has not been investigated in this exploratory analysis. See Sec.~\ref{sec:caveats}.

Note that all these features of the predicted gamma-ray flux
distribution, both with and without dark matter, fit just on the edge of
the experimental uncertainty on this mean value ($6 \pm 1\times 10^{-10}
~\SAunit$, Ref.~\cite{Abdo2010spectrum}), and may be open to exploitation by one-point function methods.

As the sum of a diffuse background and a few unresolved sources, the
gamma-ray background in Eq.~\eqref{eq:EGBPF} has the same origin as the
dark matter background.
By adopting a wider variance for the Gaussian in Eq.~\eqref{eq:EGBPF}, even the `Gaussian with a
power-law tail' form of the dark matter component can even be reproduced, corroborating our
interpretation of these features of the dark matter $P(F)$ and justifying our
faint/bright $F_*$ analysis. One might worry that this `truly diffuse plus unresolved point sources' form also undermines the
prospects of an unambiguous dark matter detection above such a background. Indeed, model-fitting one-point techniques such as that described in
Ref.~\cite{MalyshevHogg2011} cannot separate degenerate models, especially
given the angular resolution limitation (illustrated in
Fig.~\ref{ShinFig}) that prevents the dark matter $P(F)$ power-law tail
from contributing significantly to the observed flux. However, one need not worry too much: the asymmetry-induced shift of the peaks of $P(F)$ may be a sufficiently distinctive feature to extract a dark matter signal from the Fermi data nonetheless.

\subsection{The photon count distribution $P(C)$ \label{sec:PC}}

\begin{figure}[!ht]
\centering
\input{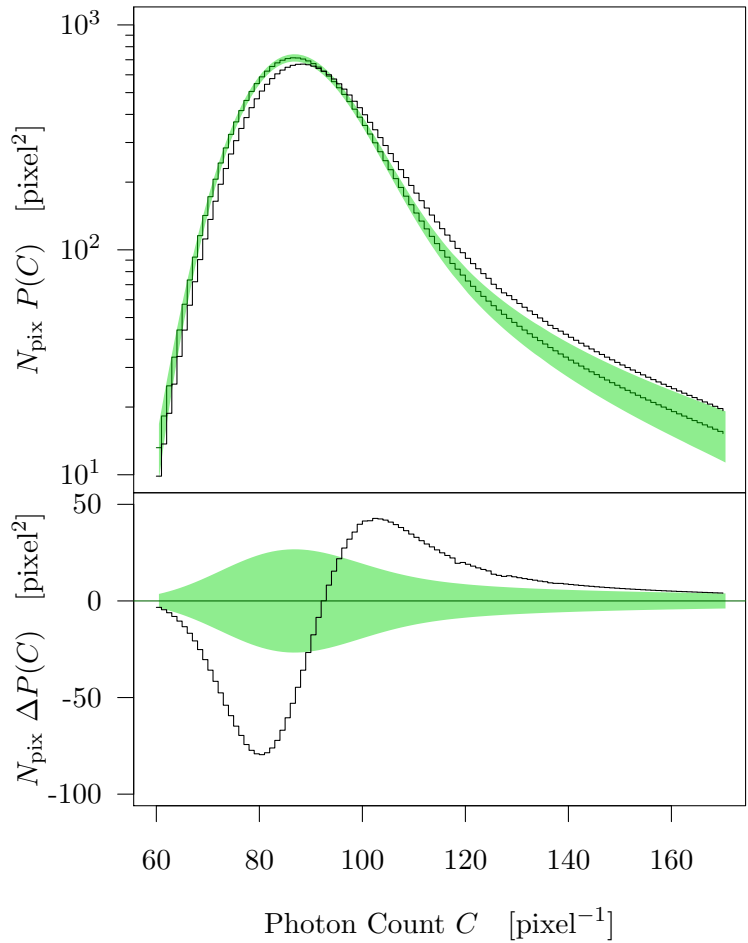}
\caption{Predicted count distributions of EBG photons with (black) and without (green) a dark matter component. The green bands represent the Poisson errors $\sigma \propto \sqrt{N\,P(C)}$ on the dark-matter-free model. The lower panel shows difference between the two models. A cross-section twice the canonical value was used to visually enhance the differences between these distributions.}
\label{fig:PC}
\end{figure}

The observable given by Fermi is not the gamma-ray flux $F$, but the discrete number of photon counts per pixel $C$. Photon arrival may then be modelled as a Poisson rate with a mean determined by the differential gamma-ray flux and the exposure  $\epsilon = (\mathrm{time}) \times (\mathrm{detector~area})\times(\mathrm{photon~energy})$. For a five-year Fermi mission, correcting for the field of view, we have an exposure of $\epsilon \approx 2.83 \times10^{14} \mathrm{~cm}^2 \mathrm{~s} \mathrm{~MeV} \mathrm{~sr} \mathrm{~pixel}^{-1}$. Marginalising over the uncertain flux distribution then gives
\begin{equation}
P(C) = \int P_\mathrm{EGB}(F) \Poisson{C}{\epsilon F} dF.
\end{equation}
This Poisson arrival uncertainty substantially smooths away the differences between the null and alternate flux models, as evidenced by Fig.~\ref{fig:PC}. However, the percent-level shift between the low-flux peaks due to the dark matter distribution's skewness survives, since a percent difference with $C \sim \mathcal{O}(100)$ is still a few photons. There is also a larger, opposite shift in the point-source-driven high-flux tail due to our imposed value of the distribution's mean.

We can define, given our number of pixels $N_\mathrm{pix}$, the test statistic \cite{lee2009Microhalo} \begin{equation}
\chi^2 = \sum_{C} \left(\frac{N_\mathrm{pix}[P_\mathrm{Null}(C)-P_\mathrm{Alt}(C)]}{\sqrt{N_\mathrm{pix} P_\mathrm{Null}(C)}}\right)^2,
\end{equation} The choice of bounds for the sum over count bins is
somewhat arbitrary; a formal optimisation of this test statistic would
be beyond the scope of this analysis. We choose to focus on
the peak $65 < C < 165$ of the distribution, in which we anticipate sufficiently many pixels per `count bin' to trust a $\chi^2$ test. The lower panel of Fig.~\ref{fig:PC}
illustrates the terms of the sum in this test statistic, from which one
may obtain the $p$-value at which data with dark matter following
$P_\mathrm{Alt}(C)$ exactly would reject our dark-matter-free null
hypothesis.

\begin{figure}[h]
\centering
\input{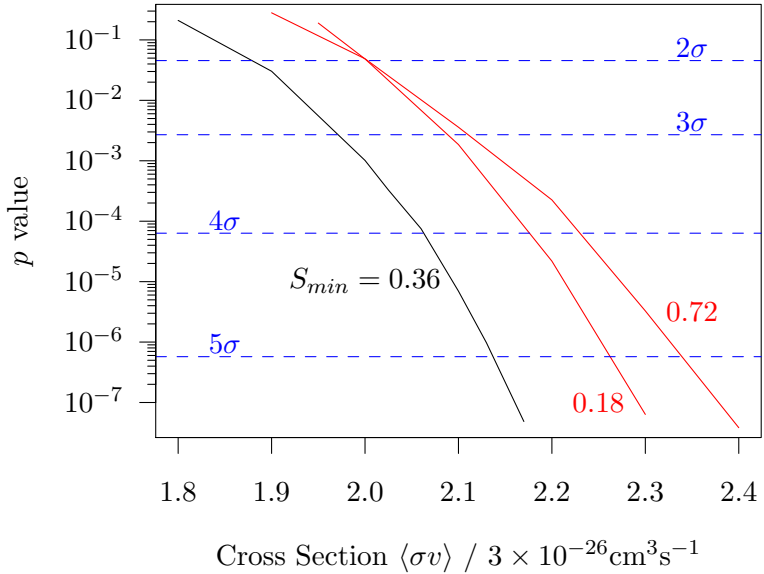}
\caption{Predicted statistical significance of a (hypothetical)
 one-point-function-only detection of a dark matter annihilation signal
 above perfectly characterised astrophysical backgrounds, as a function
 of the dark matter cross section. Curves are labelled by the flux $S_{min}$ down to which the blazar distribution is extrapolated (see Table \ref{tab:blaz}). Horizontal lines (blue, dashed)
 represent some common choices of confidence level. Including the
 energy-dependence of the flux distributions would improve these
 results, at the cost of a greater dependence on the annihilation
 spectrum.}

\label{fig:pvalues}
\end{figure}

For the fiducial dark matter model, the percent-level shift between these peaks is just small enough that the null cannot be rejected by the data. Since the
dark matter cross section $\langle \sigma v \rangle$ enters in our model
only as a proportionality factor for the dark matter halo luminosity, we
can rescale our $P_\mathrm{DM}(F)$ to inexpensively repeat this
detectability study for higher values of the cross section. We can then
forecast for which values of the cross section a dark matter component
would become distinguishable from the background using the one-point
function alone. This is summarised in Fig.~\ref{fig:pvalues}, which
shows that (given a perfect understanding of the backgrounds) the
one-point function could probe a dark matter annihilation signal with a
cross section roughly a factor two times larger than the
canonical value $3\times 10^{-26}~\mathrm{cm}^{3}\,\mathrm{s}^{-1}$.
Fig.~\ref{fig:pvalues} also shows that our fiducial choice of lower blazar flux extrapolation is fortuitously close to a detectability optimum, but our forecast does not deteriorate much upon rescaling this value.

This result, complementary to two-point function analyses, could even be
strengthened by including the energy dependence of the differential
flux to break the degeneracy with the astrophysical backgrounds
\cite{carlson2014improving}. Such a study would remain sensitive to (but would allow a quantitative analysis of) the assumptions and
uncertainties of the astrophysical background model. Yet, even without
this spectral input, our forecasted one-point upper limit on the cross-section is on par with the most recent (spectral)
constraints~\cite{Ajello:2015mfa} based on the mean value alone.

In addition to the extragalactic dark matter flux, there will be a
component due to Galactic substructures.
The one-point distribution of such a Galactic component has been
predicted~\cite{lee2009Microhalo}, and similarly features a power-law
high-flux tail.
Due to the energy spectrum Eq.~\eqref{eq:partphys}, if the mean intensity from
subhalos at the anticenter integrated above 10~GeV  is $\sim 10^{-10}
~\mathrm{cm}^{-2}~\mathrm{s}^{-2}~\mathrm{sr}^{-1}$~\cite{lee2009Microhalo},
then the mean differential intensity at 1~GeV is $\langle I \rangle \sim
10^{-12} ~\SAunit$.
This is of the same order of magnitude as the extragalactic component
discussed above, and with the same high-flux $F^{-2.5}$ power-law tail.
Thus, including the Galactic component would further enhance the
expected signal-to-noise for potential detection.
We finally note that the Galactic component will show a dipole feature,
with more flux from the Galactic center than the anticenter, which can in
principle be used to discriminate it from the isotropic extragalactic
component.

\subsection{Caveats \label{sec:caveats}}

There are a number of caveats on the results presented in this study.
Firstly, a large number of assumptions were used to simplify the
hierarchical model without a proper sensitivity analysis.
We assumed there is no scatter in
the halo parameters [Eq.~\eqref{eq:halo-model-delta-functions}], or any uncertainty on the average number $N^\prime$ of halos in each
pixel.
An NFW profile was assumed for the halos despite the fact that Einasto
and uncusped profiles would probably give less flux.
The uncertainty incurred by extrapolating the mass function $dN/dM$ and
the boost models $B(M)$ down to $M_\mathrm{min}$, is compounded by our
ignorance of the actual value of $M_\mathrm{min}$.

Secondly, we have not studied how our results depend on pixel size,
particularly the effects of source extension. We have merely assumed (Sec.~\ref{ptsrc}) that all sources are point-like, since there are on average only 0.28 extended dark matter sources per pixel, a negligible fraction of all $N^\prime = 7\times 10^{21}$
halos (Sec.~\ref{extsourceperpixel}).
However, extended sources must be either
massive or nearby (Sec.~\ref{P1SA}), and therefore would tend to have
large fluxes, affecting the distribution $P_1^{F>F_*}(F)$. Our point-source-based $P_1(F)$ is therefore not applicable to these objects at high flux. That this compromises the analysis should be obvious from
Fig.~\ref{fig:clusters}: the one-point functions do not account for the
clusters' extension, and the clusters' fluxes clearly do not live in the
domain of the PDF they should be drawn from.
Thus, extended sources should be dealt with in a complete analysis.
Some of the elements of such an analysis (such as an energy-dependent
angular resolution, a substructure-boost-dependent mass-to-solid-angle
conversion, or a redshift-dependent mass threshold $M_\mathrm{Ext}(z)$)
have been presented in the main text. Note, however, that the distribution of faint and distant sources $P_1^{F<F_*}(F)$ used to derive $\mathcal{G}_\mathrm{CLT}^{F<F_*}(F)$ is not compromised by the extension of bright and nearby sources; as long as the faint/bright split in the method of Sec.~\ref{sec:CLTMC} excludes faint extended sources (which can be guaranteed by a suitable choice of $F_*$), only the Monte Carlo based on bright sources would need be revised.

Thirdly, we have not considered the energy spectrum of our dark matter
annihilation signal, besides noting that it is a very relevant quantity
in Fig.~\ref{fig:P1SA} (which is clearly insufficient).
Using differential fluxes throughout this study is a first step in this
direction.
On one hand, the choice of the particle dark matter spectrum $dN/dE$ is
mostly unrelated to structure formation and only contributes an
energy-dependent normalisation which, from a particle physics
perspective, is almost completely arbitrary.
Hence, it would not have much predictive power and matter in practice
only for data fitting.
On the other hand, at higher energies the gamma-ray absorption matters,
and a more sophisticated model than Eq.~\eqref{eq:absorption} for this quantity
would be required. The energy resolution of the instrument ($\Delta E / E = 9\%$ at 1~GeV~\cite{Fermi2009Specs}) would also need to be accounted for in the model.

Finally, our Gaussian model for the non-blazar isotropic components of the EGB is clearly inadequate (and even the blazar model is somewhat simplistic), since rigorously accounting for all the astrophysics would require an entirely separate analysis.
Consequently the one-point functions and forecasted limits that depend on this input must be understood as exploratory and methodologically illustrative.

\section{Conclusions \label{sec:concl}}

We constructed a hierarchical model that predicts, using analytical
models of $\Lambda$CDM structure formation, the flux distribution of
gamma rays from extragalactic dark matter annihilation in unresolved
point sources.
The uncertainties on this flux subject to the modeling choices we
studied are typically percent-level; in the case of the substructure
boost function, they remain smaller than a factor of three.
We then compute, without requiring any additional physical assumptions,
the flux distribution per pixel $P(F)$, which has the characteristic
form of an isotropic diffuse Gaussian matched at high flux to the
point-source distribution with a power-law slope of $-2.5$.
This distribution is non-Gaussian and asymmetric; however the most likely flux and the mean flux are comparable at the percent-level in all but the optimistic boost model, salvaging
previous `mean intensity' constraints on the dark matter properties from this potential systematic effect.

The fluxes predicted for our fiducial model lie just within the reach of
the Fermi-LAT, and should be observable by the tenth year of the
mission.
We also showed that the distinctive features of the power-law-tailed
Gaussian distribution all live above Fermi's angular resolution.
Therefore, the extragalactic gamma-ray emission due to dark matter
annihilation constitutes an irreducible and significant background for
point-source annihilation searches with clusters or dwarf spheroidals.
Ironically, an optimistic boost model would be detrimental to these searches, by deteriorating the signal-to-noise of these point sources (to unity or worse for galaxy clusters).

We also discussed the astrophysical backgrounds from which a dark matter
annihilation signal would need to be extracted.
These include unresolved blazars (which contribute an order of magnitude
more flux than the fiducial dark matter model) and other
diffuse components, which were all convolved together into a total model
for the gamma-ray background.
The scarcity of unresolved blazars make this distribution quite rich in
features; most prominently, it has two distinct peaks of most probable
fluxes, the inter-peak gap being very sensitive to the dark matter
component.

Even accounting for the Poisson noise of photon arrivals that come with
such low fluxes, a contribution to the gamma-ray
background of the order of a vanilla WIMP model may be detectable above well-characterised astrophysical
backgrounds using the flux distribution alone.
Using the energy-dependence of the flux distribution should further
break the degeneracy between the components of the gamma-ray background, and should allow
one-point function methods to complement and strengthen existing
constraints set by two-point-function analyses.


\begin{acknowledgments}
 This work was supported by the Netherlands Organisation for Scientific
 Research through Vidi grant (MRF and SA).
\end{acknowledgments}

\appendix

\section{Failure of the central limit theorem\label{CLTFAIL}}

When constructing $P(F)$ from many individual sources (each an independent and identically distributed realisation of $P_1(F)$), it is easy to show that we expect significant deviations from a Normal distribution, even
for the large number $N^\prime \approx 7\times10^{21}$ of sources per pixel.
The ratio between the mean flux and the standard deviation of the fiducial model
is approximately $(\mu/\sigma)_{N^\prime} = 0.23$, which (using the
Gaussian's associated cumulative distribution) would firmly place 40\%
of a Gaussian $P(F)$ at negative fluxes.
This is not only mathematically impossible (given that the sum of
positive random variables \emph{must} be positive), it is also
physically nonsensical.
Note that using distributions with widths instead of delta functions for
$P(N^\prime)$ or \Poisson{k}{N^\prime} makes this problem even worse by
broadening the unphysical `Gaussian' $P_k$.

We can also show that we expect deviations from a Gaussian with slightly
more rigour:
The $k$-autoconvolution definition of $P_k(F)=(P_1)^{\star
k}(F)$ prompts us to work in the Fourier space of probability
distributions, i.e., with the characteristic functions $\phi_1(t)$ and
$\phi_k = (\phi_1)^k$.
Marginalising $P_k(F)$ with the Poisson distribution
\Poisson{k}{N^\prime} gives~\cite{scheuer1957statistical}:
\begin{eqnarray}
P(F|{N^\prime}) &=& \sum_{k=0}^\infty \frac{e^{-{N^\prime}} {N^\prime}^k}{k!} \mathcal{F}^{-1}\left[\phi_1^k\right]\\
\nonumber &=& \mathcal{F}^{-1}\left[e^{-{N^\prime}}\sum_{k=0}^\infty\frac{(\mu\phi_1)^k}{k!}\right] \\
\nonumber &=&\mathcal{F}^{-1}\left[e^{{N^\prime}(\phi_1-1)}\right]\,,
\end{eqnarray}
where we have used linearity of the (inverse) Fourier transform and have
recognised the power series expansion of the exponential function.
By Taylor expanding the characteristic function $\phi_1$, we generate
(by construction) the first moments of $P_1$:
\begin{equation}
\phi_1 = 1 + \mathbb{E}(P_1) (it) + \mathbb{V}(P_1)\frac{(it)^2}{2} +
 \mathrm{o}(-it^3)\,,
\end{equation}
such that the characteristic function associated to $P(F|\mu)$ becomes
(to second order in $t$)
\begin{equation}
\phi_{P(F|{N^\prime})}= \exp\left(\left[{N^\prime}\mathbb{E}(P_1) \right](it) - \left[{N^\prime}\mathbb{V}(P_1)\right]\frac{t^2}{2}\right) e^{{N^\prime} \mathrm{o}(-it^3)}\,.
\end{equation}
We recognise this first exponential as the characteristic function of a
Normal distribution, so
\begin{equation}
P(F|{N^\prime}) \approx \Gaussian{F}{{N^\prime}\mathbb{E}(P_1), {N^\prime}\mathbb{V}(P_1)} \star \mathcal{F}^{-1}\left[e^{{N^\prime}\mathrm{o}(-it^3)}\right]\,.
\end{equation}
If $P({N^\prime})$ is a $\delta$-function we recover upon marginalisation the near-Gaussian
form of $P(F)$ anticipated in the main text, with explicit deviations
due to higher moments convolved in. Proving that these deviations lead to a power-law tail at high flux is of less interest than the observation, in the main text, that in this regime the flux is almost completely due to a single bright source. This leads, in the next section, to a Monte Monte CarloCarlo analysis of these corrections.

The analysis above does not assume ${N^\prime},k
\rightarrow \infty$.
This reflects the fact that we have \emph{not} derived the Central limit Theorem (CLT), the
`Gaussian $\star$ Corrections' form was derived for a finite number of
sources. However, the CLT is still valid
asymptotically as $k \rightarrow \infty$ since $P_1$ does have a finite
variance despite its power-law-like behaviour.

\section{Our method \label{methodappendix}}

\subsection{Rationale of our method}

The deviations from Gaussianity in the sum $P_k$ of random variables are
bounded by the Berry-Esseen (BE) theorem~\cite{petrov1975Sums,
petrov1995LimitTheorems}, which relates the Kolmogorov-Smirnov distance
(the largest deviation at any point), to the `absolute
skewness'\footnote{Formally, $\rho_1$ is the sum of the partial
skewnesses above and below the mean. We simply call $\rho_1$ the
skewness.} of $P_1(F)$ and the large but finite number of dark matter halos $k$:
\begin{equation}
\sup_F \left|\mathrm{CDF}_k(F) - \Phi(F)\right| \le \frac{C
 \rho_1}{\sqrt{k}}\quad,\quad \rho_1 =
 \frac{\mathbb{E}(|X-\mu_1|^3)}{\sigma_1^3}\,,
\end{equation}
where $C\sim 0.5$, and where ${\rm CDF}_k$ and $\Phi(F)$ are the cumulative
distribution for $P_k$ and the Normal distribution, respectively.
We note that the absolute value in $\rho_1$ distinguishes it from the
conventional skewness $\gamma_1$, and gives the inequality $\rho_1 \ge
\gamma_1$. For our power-law-like $P_1(F)$, the skewness is huge and the BE bound
$C\rho_1 k^{-1/2}$  is uninformative: the Kolmogorov-Smirnov distance (a
quantity definitionally less than one), is `constrained' by the first
few moments of our fiducial $P_1(F)$ to be less that about $10^8$.

This perspective suggests splitting $P_1(F)$ into low-flux and high-flux
contributions, to reduce the skewness of each contribution.
In the following, we (i) derive and (ii) cross-check the Monte Carlo
method presented in the main text.

\subsection{Deriving the form of $P_k(F)$  \label{binomialPoisson}}

We derive the behaviour for $P_k$ as follows: marginalise $P_1(F)$ into
high and low flux contributions,
\begin{equation}
P_1(F) = (1-\epsilon) P_1^{F<F_*}(F) + \epsilon P_1^{F>F_*}(F)\,,
\end{equation}
where $P_1^{F<F_*, F>F_*}(F)$ are the normalised distributions of fluxes
below and above the truncation flux $F_*$ (see Fig.~\ref{fig:Fstarscheme} of the main text), and the fraction of high-flux
sources,
\begin{equation}
 \epsilon = \epsilon(F_*) = \int_{F_*}^\infty P_1(F) dF\,,
\end{equation}
is the relative normalisation of these two distributions.
Clearly the smaller we choose $F_*$, the larger $\epsilon$ becomes.
The number of faint sources $k_*$ is then simply $k(1-\epsilon)$.

$P_k$ is the $k$-autoconvolution of the sum above: we may then write $P_k$
as the sum
\begin{equation}
\label{eq:reallyugly}
P_k = \sum_{i=0}^k \frac{k!}{i!(k-i)!} \left(1-\epsilon \right)^{k-i}
\epsilon^i \left[P_{k-i}^{F<F_*} \star P_i^{F>F_*}\right]\,.
\end{equation}
This can easily be shown by induction or by using the binomial theorem
on the associated characteristic function
\begin{equation}
\phi_k=(\phi_1)^k=((1-\epsilon) \phi_1^{F<F_*} + \epsilon
 \phi_1^{F>F_*})^k\,.
\end{equation}

Unless $k\gg i$, the contribution of the $i$th term is vanishingly
small.
Physically, this reflects the low probability of having many sources
from the high-flux tail of the `power-law' $P_1(F)$.
The low-flux autoconvolution $P_{k-i}^{F<F_*}$ is very nearly Gaussian
when $F_*$ is small (i.e. when the variance of $P_{1}^{F<F_*}$ is
small).
Each term of the sum is then an isotropic background Gaussian, convolved
with high fluxes from point sources (drawn from $P_1^{F>F_*}$).
But, since $k \gg i$ the low-flux contribution $P_{k\gg i}^{F<F_*}$
factorizes (see Eqn.~\eqref{eq:endpoint} below).

How many terms do we need to consider for the gargantuan sum
[Eq.~\eqref{eq:reallyugly}] to converge?
We expect a number $k \epsilon$ of high flux sources on average.
In other words, the number $i$ of high flux sources in a pixel is drawn
from the Poisson distribution  \Poisson{i}{k\epsilon}.
This is reflected by the binomial expansion: for $1\gg\epsilon$ (which also implies
$k\gg i$) we have
\begin{equation}
\frac{k!}{i!(k-i)!} \left(1-\epsilon \right)^{k-i} \epsilon^i \approx \frac{(k\epsilon)^i}{i!} = e^{k\epsilon}\, \Poisson{i}{k\epsilon},
\end{equation}
such that (reassembling all the elements of the discussion above)
\begin{equation}
 P_k \approx P_{k}^{F<F_*} \star e^{k\epsilon}
  \left[ \sum_{i=0}^k \Poisson{i}{k\epsilon} P_i^{F>F_*}\right]\,.
  \label{eq:endpoint}
\end{equation}
Therefore, the physically relevant terms in the sum are those in the
5$\sigma$ band $k\epsilon-5\sqrt{k\epsilon}<i<k\epsilon +
5\sqrt{k\epsilon}$ of the distribution $\Poisson{i}{k\epsilon}$.

The bracketed sum, even with its large number $k$ of terms, can be
evaluated by Monte Carlo: for each realisation, draw $i$ from a Poisson
distribution, then draw that many fluxes from $P_1^{F>F_*}$, and sum
over the fluxes.
The histogram of $F_{\rm tot}^{F>F_*}$ over many realisations then
approximates the bracketed term ($i$ is marginalised automatically by
the algorithm).
We then rescale by a constant $e^{k\epsilon}$ (absorbed into the normalisation of
$P_k(F)$) and convolve with the isotropic background of low-flux sources
(just a thin delta-function in practice) to get $P_k$.

\subsection{Cross checks}

We should, of course, check that our choice of $F_*$ is physically
sensible.
If we want to split $P_k(F)$ into an isotropic background and candidate
point sources, we must verify that our faint sources do not contribute
more than zero or one photons each.
Recall that a single GeV photon per pixel in 5 years of Fermi data
corresponds to a differential flux/intensity of $F = 6 \times 10^{-6} \; \myunit$.
Choosing $F_* \approx 1.1 \times 10^{-8} \; \myunit$, over two orders of magnitude smaller, guarantees that
faint sources are indeed faint. This value of $F_*$ was in fact chosen algorithmically, such that the BE bound for $P^{F<F_*}(F)$ be $C\rho k^{-1/2}=0.005$. The applicability of the CLT in this low-flux component is thereby justified.

\begin{figure}[t]
\centering
\input{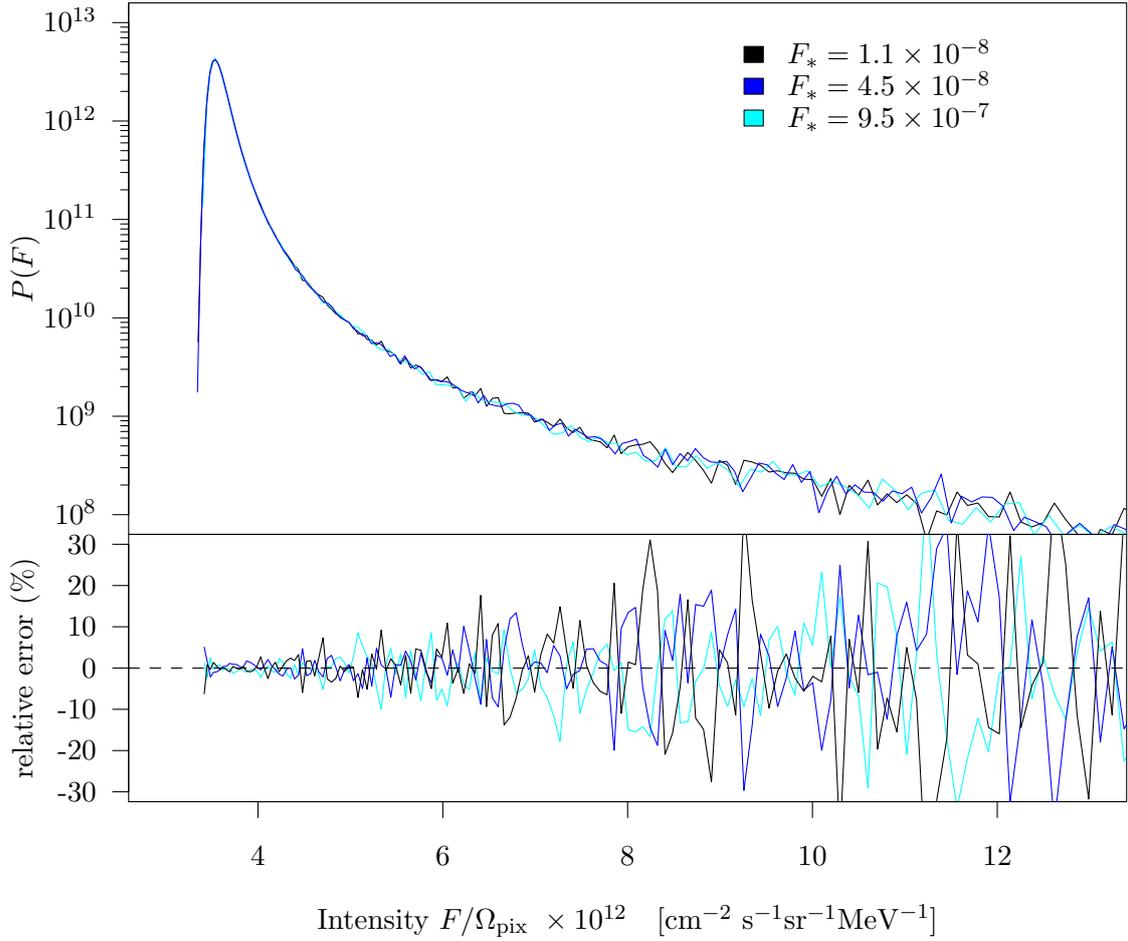}
\caption{Local sensitivity analysis of our Monte Carlo on the cutoff flux $F_*$, with $10^6$ samples of
 $P_{k_*}^{F>F_*}$ for each $F_*$. The values of the cutoff $F_*$ quoted in the legend are in units of $\myunit$. The error in the lower panel is
 relative to the average of the three Monte Carlo simulations.}
\label{fig:FstarSA}
\end{figure}

We must also check that our choice of $F_*$ does not noticeably
influence the final distribution $P(F)$: This cut is a mathematical
artifact.
Nature does not fundamentally distinguish halos that appear `faint' or
`bright' to a specific observer, with an arbitrary $F_*$.
In Fig.~\ref{fig:FstarSA} we present a (local) sensitivity analysis using
three different $F_*$ cuts, which shows that even a cutoff two orders of magnitude greater than ours (and just below the one-photon-per-five-years threshold) would have given satisfactory results (although the BE bound on the low-flux dark matter component would have been appreciably weaker).
The relative error near the peak of the distribution is percent-level
despite order-of-magnitude variations, even for a relatively small Monte
Carlo ($10^6$ realisations).
This error has an expectation consistent with zero and scales with the
size of the Monte Carlo, as expected of an Monte Carlo sampling noise.

\begin{figure}[h]
\centering
\input{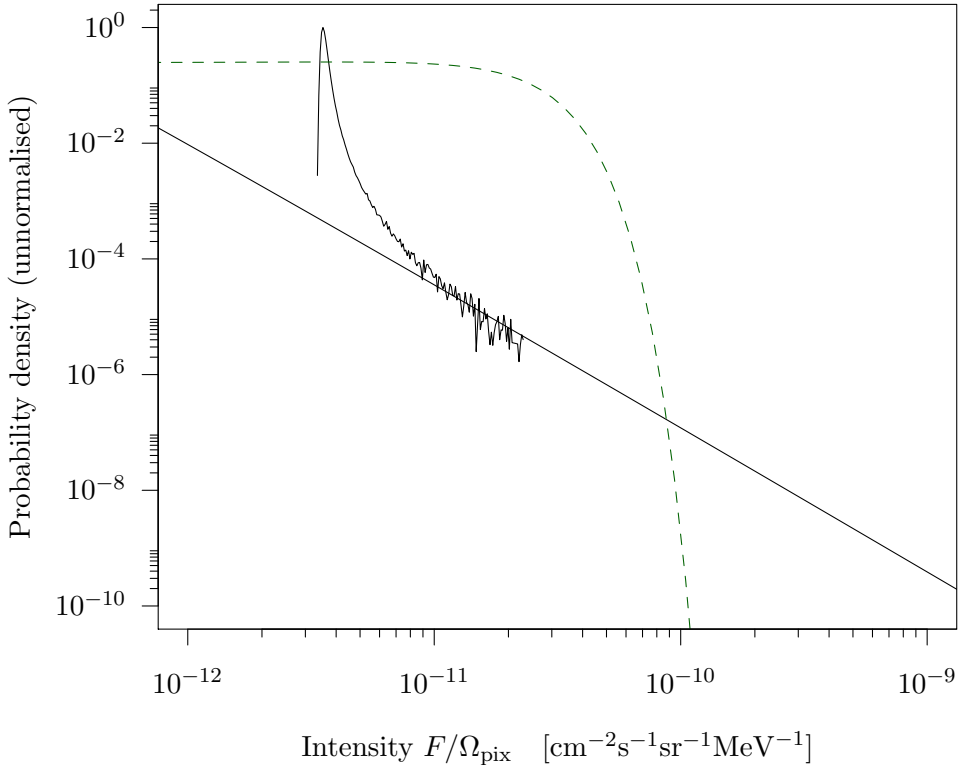}
\caption{Unnormalised $P(F)$ and $P_1(F)$ (solid, black) for the
 fiducial model. The scaling factor to achieve the matching between the
 two distributions is not a free parameter, it is the number of halos
 $k-k_* = 1864$ contributing to the high-flux tail in the fiducial
 model. We show also the (unnormalised) Gaussian with equivalent
 mean and variance predicted by a naive CLT (green, dashed).}
\label{fig:Matching}
\end{figure}

Although $P(F)$ is increasingly poorly-sampled by our Monte Carlo
(Fig.~\ref{fig:FstarSA}) as the flux increases, the behaviour of this
tail remains well-known: since the probably of a high-flux source is
relatively low, the tail of $P(F)$ should look like $P_1(F)$.
This behaviour is illustrated in Fig.~\ref{fig:Matching}.
The power-law tail continues until its cutoff (Fig.~\ref{fig:P1SA}), so
the power-law tail of $P(F)$ spans roughly ten orders of magnitude in
flux.
In light of the Gaussian intuition of exponential tails, this power-law
tail explains how our $P(F)$ can support such a large variance despite
the sharpness of its peak.

\section{Angular resolution limit \label{sec:Binomial}}

We can compute the probability of seeing a dark matter signal above a
given flux, by looking at the exceedance (complementary
cumulative) distribution $\Psi(F)$ associated to $P(F)$:\[
\Psi(F_\mathrm{max}) = \int_{F_\mathrm{max}}^\infty P(F) dF.
\]The probability of not realising this high flux tail in $N_\mathrm{pixel}$ trials is then given by the binomial distribution, and we want to solve the following for $F_\mathrm{max}$: \begin{equation}
\mathcal{B}(0|N_\mathrm{pixel},\Psi(F_\mathrm{max})) \le \alpha \quad,\quad \alpha=\left\{0.05,0.01,0.001\right\}.
\end{equation} For $k=0$ successes and $p=\Psi(F_\mathrm{max}) \ll 1$, we can expand this in a binomial series that we can truncate at first order when $\Psi(F_\mathrm{max}) \ll N_\mathrm{pixel}^{-1}$: \begin{equation}
1-N_\mathrm{pixel} \Psi(F_\mathrm{max}) + \mathcal{O}(N_\mathrm{pixel} \Psi(F_\mathrm{max}) )^2 \le \alpha.
\end{equation} When $F_{\rm max}$ lies in the power-law tail of $P(F)$, we can use the high-flux-tail equivalence (Fig.~\ref{fig:Matching}) of $P(F)$ and $P_1(F)$ to find, approximately, \begin{equation}
\int_{F_\mathrm{max}}^\infty P_1(F) dF \le \frac{1-\alpha}{N_\mathrm{pixel}}~.
\end{equation}
For a power-law-like $P_1(F) \approx A F^\gamma, \gamma<0$ this integration yields\begin{equation}
\int_{F_\mathrm{max}}^\infty A F^\gamma dF = - A (F_\mathrm{max})^{\gamma+1} / (\gamma+1) = - F_\mathrm{max} P_1(F_\mathrm{max}) /  (\gamma+1).
\end{equation}
In the limit that $(\alpha,\gamma) \to (0,-2.5)$ we reproduce Eq.~\eqref{eq:angresfmax} from the main text. More generally, for $\alpha \to 0$ and $\gamma=-2.5+\delta\gamma$, we have
\begin{equation}
F_\mathrm{max} P_1(F_\mathrm{max}) \le \frac{1.5}{N_\mathrm{pixel}} - \frac{\delta\gamma}{N_\mathrm{pixel}}~.
\end{equation} Since (as visible in Fig \ref{fig:P1SA}) the highest fluxes in the tail of $P_1(F)$ have a steeper log-slope than $-2.5$ (i.e. $\delta\gamma<0$), the angular resolution limit of Eq.~\eqref{eq:angresfmax} is actually more optimistic than would be warranted by a more precise calculation.

\bibliographystyle{JHEP}
\bibliography{Bib}

\providecommand{\href}[2]{#2}\begingroup\raggedright\begin{thebibliography}{10}

\bibitem{Fermi2009Specs}
{\bf Fermi-LAT} Collaboration, W.~Atwood et~al., {\it {The Large Area Telescope
  on the Fermi Gamma-ray Space Telescope Mission}},  {\em Astrophys.\ J.} {\bf
  697} (2009) 1071--1102, [\href{http://arxiv.org/abs/0902.1089}{{\tt
  arXiv:0902.1089}}].

\bibitem{Ackermann:2014usa}
{\bf Fermi-LAT} Collaboration, M.~Ackermann et~al., {\it {The spectrum of
  isotropic diffuse gamma-ray emission between 100 MeV and 820 GeV}},  {\em
  Astrophys.\ J.} {\bf 799} (2015), no.~1 86,
  [\href{http://arxiv.org/abs/1410.3696}{{\tt arXiv:1410.3696}}].

\bibitem{Ackermann:2012uf}
{\bf Fermi-LAT} Collaboration, M.~Ackermann et~al., {\it {Anisotropies in the
  diffuse gamma-ray background measured by the Fermi-LAT}},  {\em Phys.\ Rev.\
  D} {\bf 85} (2012) 083007, [\href{http://arxiv.org/abs/1202.2856}{{\tt
  arXiv:1202.2856}}].

\bibitem{abdo2010fermi}
{\bf Fermi-LAT} Collaboration, {\it {The Fermi-LAT high-latitude Survey: Source
  Count Distributions and the Origin of the Extragalactic Diffuse Background}},
   {\em Astrophys.\ J.} {\bf 720} (2010) 435--453,
  [\href{http://arxiv.org/abs/1003.0895}{{\tt arXiv:1003.0895}}].

\bibitem{Ajello:2015mfa}
M.~Ajello, D.~Gasparrini, M.~S{\'a}nchez-Conde, G.~Zaharijas, M.~Gustafsson,
  et~al., {\it {The Origin of the Extragalactic Gamma-Ray Background and
  Implications for Dark-Matter Annihilation}},  {\em Astrophys.\ J.} {\bf 800}
  (2015), no.~2 L27, [\href{http://arxiv.org/abs/1501.05301}{{\tt
  arXiv:1501.05301}}].

\bibitem{Tamborra:2014xia}
I.~Tamborra, S.~Ando, and K.~Murase, {\it {Star-forming galaxies as the origin
  of diffuse high-energy backgrounds: Gamma-ray and neutrino connections, and
  implications for starburst history}},  {\em JCAP} {\bf 1409} (2014), no.~09
  043, [\href{http://arxiv.org/abs/1404.1189}{{\tt arXiv:1404.1189}}].

\bibitem{DiMauro:2013xta}
M.~Di~Mauro, F.~Calore, F.~Donato, M.~Ajello, and L.~Latronico, {\it {Diffuse
  $\gamma$-ray emission from misaligned active galactic nuclei}},  {\em
  Astrophys.\ J.} {\bf 780} (2014) 161,
  [\href{http://arxiv.org/abs/1304.0908}{{\tt arXiv:1304.0908}}].

\bibitem{DiMauro:2015tfa}
M.~Di~Mauro and F.~Donato, {\it {The composition of the Fermi-LAT IGRB
  intensity: emission from extragalactic point sources and dark matter
  annihilations}},  \href{http://arxiv.org/abs/1501.05316}{{\tt
  arXiv:1501.05316}}.

\bibitem{Cuoco:2012yf}
A.~Cuoco, E.~Komatsu, and J.~Siegal-Gaskins, {\it {Joint anisotropy and source
  count constraints on the contribution of blazars to the diffuse gamma-ray
  background}},  {\em Phys.\ Rev.} {\bf D86} (2012) 063004,
  [\href{http://arxiv.org/abs/1202.5309}{{\tt arXiv:1202.5309}}].

\bibitem{Jungman:1995df}
G.~Jungman, M.~Kamionkowski, and K.~Griest, {\it {Supersymmetric dark matter}},
   {\em Phys.\ Rept.} {\bf 267} (1996) 195--373,
  [\href{http://arxiv.org/abs/hep-ph/9506380}{{\tt hep-ph/9506380}}].

\bibitem{Hooper:2007qk}
D.~Hooper and S.~Profumo, {\it {Dark matter and collider phenomenology of
  universal extra dimensions}},  {\em Phys.\ Rept.} {\bf 453} (2007) 29--115,
  [\href{http://arxiv.org/abs/hep-ph/0701197}{{\tt hep-ph/0701197}}].

\bibitem{Fornasa:2015qua}
M.~Fornasa and M.~A. Sanchez-Conde, {\it {The nature of the Diffuse Gamma-Ray
  Background}},  \href{http://arxiv.org/abs/1502.02866}{{\tt
  arXiv:1502.02866}}.

\bibitem{Ando:2015qda}
S.~Ando and K.~Ishiwata, {\it {Constraints on decaying dark matter from the
  extragalactic gamma-ray background}},
  \href{http://arxiv.org/abs/1502.02007}{{\tt arXiv:1502.02007}}.

\bibitem{AndoCrossCorrel}
S.~Ando, A.~Benoit-L{\'e}vy, and E.~Komatsu, {\it {Mapping dark matter in the
  gamma-ray sky with galaxy catalogs}},  {\em Phys.\ Rev.\ D} {\bf 90} (2014),
  no.~2 023514, [\href{http://arxiv.org/abs/1312.4403}{{\tt arXiv:1312.4403}}].

\bibitem{Ando:2014aoa}
S.~Ando, {\it {Power spectrum tomography of dark matter annihilation with local
  galaxy distribution}},  {\em JCAP} {\bf 1410} (2014), no.~10 061,
  [\href{http://arxiv.org/abs/1407.8502}{{\tt arXiv:1407.8502}}].

\bibitem{Camera:2012cj}
S.~Camera, M.~Fornasa, N.~Fornengo, and M.~Regis, {\it {A Novel Approach in the
  Weakly Interacting Massive Particle Quest: Cross-correlation of Gamma-Ray
  Anisotropies and Cosmic Shear}},  {\em Astrophys.\ J.} {\bf 771} (2013) L5,
  [\href{http://arxiv.org/abs/1212.5018}{{\tt arXiv:1212.5018}}].

\bibitem{Camera:2014rja}
S.~Camera, M.~Fornasa, N.~Fornengo, and M.~Regis, {\it {Tomographic-spectral
  approach for dark matter detection in the cross-correlation between cosmic
  shear and diffuse gamma-ray emission}},
  \href{http://arxiv.org/abs/1411.4651}{{\tt arXiv:1411.4651}}.

\bibitem{xia2011cross}
J.-Q. Xia, A.~Cuoco, E.~Branchini, M.~Fornasa, and M.~Viel, {\it {A
  cross-correlation study of the Fermi-LAT $\gamma$-ray diffuse extragalactic
  signal}},  {\em Mon.Not.Roy.Astron.Soc.} {\bf 416} (2011) 2247--2264,
  [\href{http://arxiv.org/abs/1103.4861}{{\tt arXiv:1103.4861}}].

\bibitem{Shirasaki:2014noa}
M.~Shirasaki, S.~Horiuchi, and N.~Yoshida, {\it {Cross-Correlation of Cosmic
  Shear and Extragalactic Gamma-ray Background: Constraints on the Dark Matter
  Annihilation Cross-Section}},  {\em Phys.\ Rev. D} {\bf 90} (2014), no.~6
  063502, [\href{http://arxiv.org/abs/1404.5503}{{\tt arXiv:1404.5503}}].

\bibitem{Fornengo:2014cya}
N.~Fornengo, L.~Perotto, M.~Regis, and S.~Camera, {\it {Evidence of
  Cross-correlation between the CMB Lensing and the $\Gamma$-ray sky}},  {\em
  Astrophys.\ J.} {\bf 802} (2015), no.~1 L1,
  [\href{http://arxiv.org/abs/1410.4997}{{\tt arXiv:1410.4997}}].

\bibitem{Xia:2015wka}
J.-Q. Xia, A.~Cuoco, E.~Branchini, and M.~Viel, {\it {Tomography of the
  Fermi-LAT gamma-ray diffuse extragalactic signal via cross-correlations with
  galaxy catalogs}},  {\em Astrophys.\ J.\ Suppl.} {\bf 217} (2015), no.~1 15,
  [\href{http://arxiv.org/abs/1503.05918}{{\tt arXiv:1503.05918}}].

\bibitem{Regis:2015zka}
M.~Regis, J.-Q. Xia, A.~Cuoco, E.~Branchini, N.~Fornengo, et~al., {\it
  {Particle dark matter searches outside the Local neighborhood}},
  \href{http://arxiv.org/abs/1503.05922}{{\tt arXiv:1503.05922}}.

\bibitem{berlin2014stringent}
A.~Berlin and D.~Hooper, {\it {Stringent Constraints on the Dark Matter
  Annihilation Cross Section From Subhalo Searches with the Fermi Gamma-Ray
  Space Telescope}},  {\em Phys.\ Rev.\ D} {\bf 89} (2014), no.~1 016014,
  [\href{http://arxiv.org/abs/1309.0525}{{\tt arXiv:1309.0525}}].

\bibitem{MalyshevHogg2011}
D.~Malyshev and D.~W. Hogg, {\it {Statistics of gamma-ray point sources below
  the Fermi detection limit}},  {\em Astrophys.\ J.} {\bf 738} (2011) 181,
  [\href{http://arxiv.org/abs/1104.0010}{{\tt arXiv:1104.0010}}].

\bibitem{lee2009Microhalo}
S.~K. Lee, S.~Ando, and M.~Kamionkowski, {\it {The Gamma-Ray-Flux Probability
  Distribution Function from Galactic Halo Substructure}},  {\em JCAP} {\bf
  0907} (2009) 007, [\href{http://arxiv.org/abs/0810.1284}{{\tt
  arXiv:0810.1284}}].

\bibitem{Dodelson:2009ih}
S.~Dodelson, A.~V. Belikov, D.~Hooper, and P.~Serpico, {\it {Identifying Dark
  Matter Annihilation Products In The Diffuse Gamma Ray Background}},  {\em
  Phys.\ Rev.\ D} {\bf 80} (2009) 083504,
  [\href{http://arxiv.org/abs/0903.2829}{{\tt arXiv:0903.2829}}].

\bibitem{Baxter:2010fr}
E.~J. Baxter, S.~Dodelson, S.~M. Koushiappas, and L.~E. Strigari, {\it
  {Constraining Dark Matter in Galactic Substructure}},  {\em Phys.\ Rev.\ D}
  {\bf 82} (2010) 123511, [\href{http://arxiv.org/abs/1006.2399}{{\tt
  arXiv:1006.2399}}].

\bibitem{carlson2014improving}
E.~Carlson, D.~Hooper, and T.~Linden, {\it {Improving the Sensitivity of
  Gamma-Ray Telescopes to Dark Matter Annihilation in Dwarf Spheroidal
  Galaxies}},  {\em Phys.Rev.} {\bf D91} (2015), no.~6 061302,
  [\href{http://arxiv.org/abs/1409.1572}{{\tt arXiv:1409.1572}}].

\bibitem{NFW}
J.~F. Navarro, C.~S. Frenk, and S.~D. White, {\it {A Universal density profile
  from hierarchical clustering}},  {\em Astrophys.\ J.} {\bf 490} (1997)
  493--508, [\href{http://arxiv.org/abs/astro-ph/9611107}{{\tt
  astro-ph/9611107}}].

\bibitem{scheuer1957statistical}
P.~A. Scheuer, {\it A statistical method for analysing observations of faint
  radio stars},  in {\em Mathematical Proceedings of the Cambridge
  Philosophical Society}, vol.~53, pp.~764--773, Cambridge Univ Press, 1957.

\bibitem{petrov1975Sums}
V.~V. Petrov, {\em Sums of independent random variables}, vol.~82.
\newblock Springer, 1975.

\bibitem{Planck2013CosmoParams}
{\bf Planck} Collaboration, P.~Ade et~al., {\it {Planck 2013 results. XVI.
  Cosmological parameters}},  {\em Astron.\ Astrophys.} {\bf 571} (2014) A16,
  [\href{http://arxiv.org/abs/1303.5076}{{\tt arXiv:1303.5076}}].

\bibitem{Green:2005fa}
A.~M. Green, S.~Hofmann, and D.~J. Schwarz, {\it {The First wimpy halos}},
  {\em JCAP} {\bf 0508} (2005) 003,
  [\href{http://arxiv.org/abs/astro-ph/0503387}{{\tt astro-ph/0503387}}].

\bibitem{Loeb:2005pm}
A.~Loeb and M.~Zaldarriaga, {\it {The Small-scale power spectrum of cold dark
  matter}},  {\em Phys.Rev.} {\bf D71} (2005) 103520,
  [\href{http://arxiv.org/abs/astro-ph/0504112}{{\tt astro-ph/0504112}}].

\bibitem{Bertschinger:2006nq}
E.~Bertschinger, {\it {The Effects of Cold Dark Matter Decoupling and Pair
  Annihilation on Cosmological Perturbations}},  {\em Phys.Rev.} {\bf D74}
  (2006) 063509, [\href{http://arxiv.org/abs/astro-ph/0607319}{{\tt
  astro-ph/0607319}}].

\bibitem{Hofmann:2001bi}
S.~Hofmann, D.~J. Schwarz, and H.~Stoecker, {\it {Damping scales of neutralino
  cold dark matter}},  {\em Phys.Rev.} {\bf D64} (2001) 083507,
  [\href{http://arxiv.org/abs/astro-ph/0104173}{{\tt astro-ph/0104173}}].

\bibitem{Profumo:2006bv}
S.~Profumo, K.~Sigurdson, and M.~Kamionkowski, {\it {What mass are the smallest
  protohalos?}},  {\em Phys.Rev.Lett.} {\bf 97} (2006) 031301,
  [\href{http://arxiv.org/abs/astro-ph/0603373}{{\tt astro-ph/0603373}}].

\bibitem{PressSchechter}
W.~H. Press and P.~Schechter, {\it {Formation of galaxies and clusters of
  galaxies by selfsimilar gravitational condensation}},  {\em Astrophys.\ J.}
  {\bf 187} (1974) 425--438.

\bibitem{BCEK}
J.~Bond, S.~Cole, G.~Efstathiou, and N.~Kaiser, {\it {Excursion set mass
  functions for hierarchical Gaussian fluctuations}},  {\em Astrophys.\ J.}
  {\bf 379} (1991) 440.

\bibitem{BBKS1986}
J.~M. Bardeen, J.~Bond, N.~Kaiser, and A.~Szalay, {\it The statistics of peaks
  of gaussian random fields},  {\em The Astrophysical Journal} {\bf 304} (1986)
  15--61.

\bibitem{Lapi2013Statistics}
A.~Lapi, P.~Salucci, and L.~Danese, {\it {Statistics of Dark Matter Halos from
  the Excursion Set Approach}},  {\em Astrophys.\ J.} {\bf 772} (2013) 85,
  [\href{http://arxiv.org/abs/1305.7382}{{\tt arXiv:1305.7382}}].

\bibitem{DelPopolo:2006gn}
A.~Del~Popolo, {\it {On the cosmological mass function theory}},  {\em Astron.\
  Rep.} {\bf 51} (2007) 709--734,
  [\href{http://arxiv.org/abs/astro-ph/0609166}{{\tt astro-ph/0609166}}].

\bibitem{SMTormen}
R.~K. Sheth, H.~Mo, and G.~Tormen, {\it {Ellipsoidal collapse and an improved
  model for the number and spatial distribution of dark matter haloes}},  {\em
  Mon.\ Not.\ Roy.\ Astron.\ Soc.} {\bf 323} (2001) 1,
  [\href{http://arxiv.org/abs/astro-ph/9907024}{{\tt astro-ph/9907024}}].

\bibitem{ShethTormen}
R.~K. Sheth and G.~Tormen, {\it {An Excursion set model of hierarchical
  clustering : Ellipsoidal collapse and the moving barrier}},  {\em Mon.\ Not.\
  Roy.\ Astron.\ Soc.} {\bf 329} (2002) 61,
  [\href{http://arxiv.org/abs/astro-ph/0105113}{{\tt astro-ph/0105113}}].

\bibitem{Ando:2009fp}
S.~Ando, {\it {Gamma-ray background anisotropy from galactic dark matter
  substructure}},  {\em Phys.\ Rev.\ D} {\bf 80} (2009) 023520,
  [\href{http://arxiv.org/abs/0903.4685}{{\tt arXiv:0903.4685}}].

\bibitem{Bullock}
J.~S. Bullock, T.~S. Kolatt, Y.~Sigad, R.~S. Somerville, A.~V. Kravtsov,
  et~al., {\it {Profiles of dark haloes. Evolution, scatter, and environment}},
   {\em Mon.\ Not.\ Roy.\ Astron.\ Soc.} {\bf 321} (2001) 559--575,
  [\href{http://arxiv.org/abs/astro-ph/9908159}{{\tt astro-ph/9908159}}].

\bibitem{HuKravtsov}
W.~Hu and A.~V. Kravtsov, {\it {Sample variance considerations for cluster
  surveys}},  {\em Astrophys.\ J.} {\bf 584} (2003) 702--715,
  [\href{http://arxiv.org/abs/astro-ph/0203169}{{\tt astro-ph/0203169}}].

\bibitem{SC2013Flattening}
M.~A. Sanchez-Conde and F.~Prada, {\it {The flattening of the
  concentration-mass relation towards low halo masses and its implications for
  the annihilation signal boost}},  {\em Mon.\ Not.\ Roy.\ Astron.\ Soc.} {\bf
  442} (2014) 2271, [\href{http://arxiv.org/abs/1312.1729}{{\tt
  arXiv:1312.1729}}].

\bibitem{gao2012MNRAS}
L.~Gao, C.~Frenk, A.~Jenkins, V.~Springel, and S.~White, {\it {Where will
  supersymmetric dark matter first be seen?}},  {\em Mon.\ Not.\ Roy.\ Astron.\
  Soc.} {\bf 419} (2012) 1721, [\href{http://arxiv.org/abs/1107.1916}{{\tt
  arXiv:1107.1916}}].

\bibitem{PradaKlypin}
F.~Prada, A.~A. Klypin, A.~J. Cuesta, J.~E. Betancort-Rijo, and J.~Primack,
  {\it {Halo concentrations in the standard LCDM cosmology}},  {\em Mon.\ Not.\
  Roy.\ Astron.\ Soc.} {\bf 428} (2012) 3018--3030,
  [\href{http://arxiv.org/abs/1104.5130}{{\tt arXiv:1104.5130}}].

\bibitem{diemer2014universal}
B.~Diemer and A.~V. Kravtsov, {\it {A universal model for halo
  concentrations}},  {\em Astrophys.\ J.} {\bf 799} (2015), no.~1 108,
  [\href{http://arxiv.org/abs/1407.4730}{{\tt arXiv:1407.4730}}].

\bibitem{bergstrom2001spectral}
L.~Bergstrom, J.~Edsjo, and P.~Ullio, {\it {Spectral gamma-ray signatures of
  cosmological dark matter annihilation}},  {\em Phys.\ Rev.\ Lett.} {\bf 87}
  (2001) 251301, [\href{http://arxiv.org/abs/astro-ph/0105048}{{\tt
  astro-ph/0105048}}].

\bibitem{zdziarski1989absorption}
A.~A. Zdziarski and R.~Svensson, {\it Absorption of x-rays and gamma rays at
  cosmological distances},  {\em The Astrophysical Journal} {\bf 344} (1989)
  551--566.

\bibitem{eisenstein1998baryonic}
D.~J. Eisenstein and W.~Hu, {\it Baryonic features in the matter transfer
  function},  {\em The Astrophysical Journal} {\bf 496} (1998), no.~2 605.

\bibitem{efstathiou1986isocurvature}
G.~Efstathiou and J.~R. Bond, {\it Isocurvature cold dark matter fluctuations},
   {\em Monthly Notices of the Royal Astronomical Society} {\bf 218} (1986),
  no.~1 103--121.

\bibitem{klypin2011dark}
A.~A. Klypin, S.~Trujillo-Gomez, and J.~Primack, {\it Dark matter halos in the
  standard cosmological model: Results from the bolshoi simulation},  {\em The
  Astrophysical Journal} {\bf 740} (2011), no.~2 102.

\bibitem{warren2006precision}
M.~S. Warren, K.~Abazajian, D.~E. Holz, and L.~Teodoro, {\it Precision
  determination of the mass function of dark matter halos},  {\em The
  Astrophysical Journal} {\bf 646} (2006), no.~2 881.

\bibitem{colin2004dwarf}
P.~Col{\'\i}n, A.~Klypin, O.~Valenzuela, and S.~Gottl{\"o}ber, {\it Dwarf dark
  matter halos},  {\em The Astrophysical Journal} {\bf 612} (2004), no.~1 50.

\bibitem{sasaki2014Statistical}
M.~Sasaki, P.~C. Clark, V.~Springel, R.~S. Klessen, and S.~C. Glover, {\it
  Statistical properties of dark matter mini-haloes at z>= 15},  {\em arXiv
  preprint arXiv:1405.4018} (2014).

\bibitem{ando2005EGB}
S.~Ando, {\it {Can dark matter annihilation dominate the extragalactic
  gamma-ray background?}},  {\em Phys.\ Rev.\ Lett.} {\bf 94} (2005) 171303,
  [\href{http://arxiv.org/abs/astro-ph/0503006}{{\tt astro-ph/0503006}}].

\bibitem{Ando:2005xg}
S.~Ando and E.~Komatsu, {\it {Anisotropy of the cosmic gamma-ray background
  from dark matter annihilation}},  {\em Phys.\ Rev.\ D} {\bf 73} (2006)
  023521, [\href{http://arxiv.org/abs/astro-ph/0512217}{{\tt
  astro-ph/0512217}}].

\bibitem{haines1988logarithmic}
G.~Haines and A.~G. Jones, {\it Logarithmic fourier transformation},  {\em
  Geophysical Journal International} {\bf 92} (1988), no.~1 171--178.

\bibitem{petrov1995LimitTheorems}
V.~V. Petrov, ``Limit theorems of probability theory. 1995.''

\bibitem{Ackermann:2015tah}
{\bf Fermi-LAT} Collaboration, M.~Ackermann et~al., {\it {Limits on Dark Matter
  Annihilation Signals from the Fermi LAT 4-year Measurement of the Isotropic
  Gamma-Ray Background}},  \href{http://arxiv.org/abs/1501.05464}{{\tt
  arXiv:1501.05464}}.

\bibitem{Abdo2010spectrum}
{\bf Fermi-LAT} Collaboration, A.~Abdo et~al., {\it {The Spectrum of the
  Isotropic Diffuse Gamma-Ray Emission Derived From First-Year Fermi Large Area
  Telescope Data}},  {\em Phys.\ Rev.\ Lett.} {\bf 104} (2010) 101101,
  [\href{http://arxiv.org/abs/1002.3603}{{\tt arXiv:1002.3603}}].

\bibitem{ando2012Fornax}
S.~Ando and D.~Nagai, {\it Fermi-lat constraints on dark matter annihilation
  cross section from observations of the fornax cluster},  {\em Journal of
  Cosmology and Astroparticle Physics} {\bf 2012} (2012), no.~07 017.

\bibitem{ZandanelComa}
F.~Zandanel and S.~Ando, {\it Constraints on diffuse gamma-ray emission from
  structure formation processes in the coma cluster},  {\em Monthly Notices of
  the Royal Astronomical Society} {\bf 440} (2014), no.~1 663--671,
  [\href{http://arxiv.org/abs/http://mnras.oxfordjournals.org/content/440/1/663.full.pdf+html}{{\tt
  http://mnras.oxfordjournals.org/content/440/1/663.full.pdf+html}}].

\bibitem{arad2004phase}
I.~Arad, A.~Dekel, and A.~Klypin, {\it Phase-space structure of dark matter
  haloes: scale-invariant probability density function driven by substructure},
   {\em Monthly Notices of the Royal Astronomical Society} {\bf 353} (2004),
  no.~1 15--29.

\bibitem{ackermann2011constraining-dark-matter}
M.~Ackermann et~al., {\it Constraining dark matter models from a combined
  analysis of milky way satellites with the fermi large area telescope},  {\em
  Physical Review Letters} {\bf 107} (2011), no.~24.

\end{thebibliography}\endgroup

\end{document}